\documentclass[preprint,12pt,authoryear]{elsarticle}

\usepackage{amsmath,amssymb,amsthm,bm,amstext}
\usepackage{mathrsfs}
\usepackage[T1]{fontenc}
\usepackage{ascmac}

\usepackage{xcolor}
\usepackage{ulem}

\usepackage{natbib}
\usepackage{siunitx}
\usepackage{commath}
\usepackage{comment}

\usepackage{here}

\usepackage{multirow}

\begin{document}

\begin{frontmatter}


\title{Convective regimes of internally heated steady thermal convection of temperature-dependent viscous fluid}


\author[inst1]{Hisashi Okuda}

\affiliation[inst1]{
            organization={Research Institute for Mathematical Sciences, Kyoto University},
            addressline={Kitashirakawa-Oiwake-chou, Sakyo-ku},
            city={Kyoto},
            postcode={606-8502},
            country={Japan}}

\author[inst1]{Shin-ichi Takehiro}

\author[inst2]{St\'ephane Labrosse}
\affiliation[inst2]{
            organization={LGLTPE, ENS de Lyon, Universit\'e de Lyon},
            addressline={46 all\'ee d'Italie},
            city={Lyon},
            postcode={69003},
            country={France}}


\begin{abstract}
We study the dynamical regimes of thermal convection with temperature-dependent viscosity driven by homogeneous internal heating. 
Two-dimensional steady-state convective solutions 
with the Frank-Kamenetskii viscosity 
are obtained by the Newton method for a number of different values of the Rayleigh number and the strength of the dependence of the viscosity on temperature.

By classifying the solutions with the top surface mobility, we find the sluggish lid regime between the mobile and stagnant lid regimes.
The solutions of the sluggish regime are characterized by a large viscosity contrast through the boundary layer below the conductive lid and a rapid increase of the Nusselt number with respect to the Rayleigh number.

For most solutions in the mobile and stagnant lid regimes, the Nusselt number is proportional to the $1/6$ power of the Rayleigh number, which can be derived by taking into account the effect of thin and strong downwelling plumes.

Time evolution calculations show that the steady solutions become unstable for large Rayleigh numbers, where additional downward plumes grow on the background convective flows. 
This can be explained by the timescale of the Rayleigh--Taylor instability for the thermal boundary layer between the conductive lid and the convective core, which is shorter than the timescale of horizontal advection by the background flows.

In addition, steady convective solutions with Arrhenius-law viscosity are calculated for several values of the parameters. 
The obtained regime diagram qualitatively agrees with that of
the Frank-Kamenetskii viscosity,
while there is a slight difference in the locations of the regime boundaries.

\end{abstract}


\begin{keyword}
Mantle convection \sep 
Sluggish lid regime \sep 
Scaling relationship \sep 
Rayleigh-Taylor instability \sep 
Planetary surfaces and interiors
\end{keyword}

\end{frontmatter}


\section{Introduction}
\label{sec_intro}
Thermal convection of temperature-dependent viscous fluids has been studied
as a fundamental model problem of mantle convection in terrestrial planets.
For convection driven by bottom heating,
which is an extension of the configuration of Rayleigh--B\'enard convection,
the linear stability analysis
\citep{Stengel1982} and the scaling analyses of finite-amplitude convection
\citep[e.g.][]{Solomatov1995} have been performed so far.
\cite{Solomatov1995}
classified the convective structures into three regimes
depending on the viscosity contrast in the fluid layer. 
When the viscosity contrast is small,
convection occurs in the whole layer
(small viscosity contrast or mobile regime).
When the viscosity contrast is large,
a conductive lid is formed just below the top surface which is not involved in the convective flow due to the low temperature and high viscosity
(stagnant lid regime).
There is an intermediate regime which is called transitional or sluggish lid regime,
where the conductive lid exists but it is horizontally mobile.
These regimes are distinguished by scaling relations
between the Nusselt and Rayleigh numbers
\citep{Solomatov1995}.

On the other hand, 
the main thermal forcing of mantle convection in the Earth is internal heating
rather than heating from the bottom
\citep{Treatise_Jaupart2015}.
Many other terrestrial planets, satellites, and exoplanets may also have internal heating sources
such as radioactive and tidal heating, or secular cooling
\citep{JainSolomatov2022}.
Previous studies on internally heated iso-viscous convection showed
the scaling relation between $Nu \sim Ra^{1/4}$
where $Nu$ is the Nusselt number and $Ra$ is the Rayleigh number defined by the internal heating rate
\citep{Geodynamics2014, Parmentier1994}. 
For temperature-dependent viscosity convection, 
\cite{Davaille1993} performed laboratory experiments
of convection driven by cooling from above, 
which is analogous to internally heated convection.
They suggested that the temperature difference between the convection core and the stagnant lid
was the temperature scale that defined the viscosity variations. 
\cite{Grasset1998}
performed numerical simulations in 2-D Cartesian box with aspect ratio of 2,
and classified convective solutions into no-lid and conductive lid regimes,
corresponding to mobile regime and stagnant lid regime of our description.
They found that the Nusselt number is proportional to the power of $1/4$ of the Rayleigh number in both regimes
by including the viscosity contrast in the coefficients and rescaling to the effective values for the convective layer
below the conductive lid.
Scaling relations in the stagnant lid regime of internal heating convection have also been studied
by numerical simulations in 2-D Cartesian boxes
\citep[e.g.][]{SolomatovMoresi2000}
and 3-D spherical shells 
\citep[e.g.][]{Reese2005, Huttig2011}.

However,
the regime classification of the structures of the internal heating convection is not fully discussed
compared to that of the bottom heating convection.
In particular, the intermediate regime between the mobile and stagnant lid regimes
has not been well characterized, even though it is relevant in the context of planetary evolution.
\cite{Huttig2011}
studied convection in a spherical shell driven by internal heating.
They found an intermediate regime
and characterized it 
as a horizontally elongated structure
where the spherical harmonic degree of the convective solutions is 1 or 2.
Bottom heating convection also suggested to become 
wider in horizontal wavelength than square convection cells obtained in the sluggish lid regime
\citep{Kameyama2000, Okuda2023}.

The sluggish lid regime of convection in a temperature-dependent-viscosity fluid
is important for geophysics.
The sluggish lid or the stagnant lid are considered to be the plates of the Earth's surface
\citep{Bercovici2000}.
During their evolution, terrestrial planets are expected to follow a path in the heat flow--temperature space that is constrained by the scaling law associated to the regime in which mantle convection operates \citep[e.g.][]{Sleep2000}. The regimes usually considered are magma ocean, plate tectonics and stagnant lid, the latter two being currently invoked for the present-day terrestrial planets in the solar system. Transitions between regimes may happen back and forth, leading to episodicity \citep[e.g.][]{SolomatovMoresi1996, FowlerOBrien1996, Sleep2000}. When a planet is far from the conditions for plate tectonics to arise, the relevant regimes in the solid state are the mobile, sluggish and stagnant lid.

Although more realistic conditions of viscosity will be required
for the reproduction of Earth-like plate tectonics, 
the three regimes essentially come from the temperature dependence of viscosity.
Understanding the sluggish lid regime and its boundaries with the stagnant lid regime and with the mobile regime in the parameter space is therefore important to understand the long-term evolution of terrestrial planets.

In this study, 
we investigate in detail the structures of thermal convection of a temperature-dependent viscous fluid
driven by internal heating for the mobile, sluggish lid, and stagnant lid regimes.
Various solutions of finite-amplitude steady-state convection are obtained
by the Newton method calculations
according to the method of \cite{Okuda2023}.
As a result we examine the convective structures and scaling relations of each regime
in order to characterize the sluggish lid regime.
Section 2 describes the model, the experimental setup, and the numerical method. 
Section 3 presents the results of the regime classification and the characterization of the solutions in each regime.
In Section 4, the scaling relations for the mobile regime and stagnant lid regime
are explained by modifying the classical boundary layer theory \citep[e.g.][]{Geodynamics2014}, 
the stability of the steady solutions,
the effect of the surface boundary condition,
and the regime diagram of Arrhenius viscosity convection are discussed.
Section 5 concludes the results of our study. 
\section{Formulation and numerical method}
\label{sec_method}

\subsection{Governing equations}
\label{subsec_eq}
Thermal convection of an incompressible fluid
is considered in a two-dimensional horizontal layer with an aspect ratio of 2.
The Prandtl number is taken as infinite and the Boussinesq approximation is assumed.
The domain is horizontally periodic,
and bounded by a zero-heat-flow boundary at the bottom
and an isothermal boundary at the top.
Both horizontal boundaries are impermeable and stress-free.
The horizontal and vertical axes are $x$ and $y$, respectively.
The viscosity $\eta$ is the Newtonian and 
depends on the dimensionless temperature $T$ as
$\eta \propto \exp\left( - \gamma T \right)$,
where $\gamma$ is the Frank-Kamenetskii parameter.
Dimensionless variables are defined using
the layer thickness $d$, 
thermal diffusion time $d^2/\kappa$, 
the viscosity at the top surface $\eta_0$, 
and the temperature scale $H d^2 / k$, 
with $\kappa$ the thermal diffusivity, 
$k$ the thermal conductivity 
and $H$ the uniform volumetric heating rate.
The equations governing velocity $(u,v)$ and temperature $T$ are \citep[e.g.][]{Christensen1984}:
\begin{equation}
  (u,v) = \left(
    - \frac{\partial \psi}{\partial y}, \frac{\partial \psi}{\partial x}
  \right),
\end{equation}
\begin{equation}
  \frac{\partial T}{\partial t}
  = \frac{ \partial (T, \psi) }{ \partial (x,y) }
  + \nabla^2 T
  + 1,
  \label{eq_energy}
\end{equation}
\begin{equation}
  \left( \frac{ \partial^2 }{ \partial x^2 } - \frac{ \partial^2 }{ \partial y^2 } \right)
  \left[ \eta(T) \left( \frac{ \partial^2 }{ \partial x^2 } - \frac{ \partial^2 }{ \partial y^2 } \right) \psi \right]
  + 4 \frac{ \partial^2 }{ \partial x \partial y }
      \left( \eta(T) \frac{ \partial^2 \psi }{ \partial x \partial y } \right)
    = - Ra_0 \frac{ \partial T }{ \partial x } , 
  \label{eq_stokes}
\end{equation}
\begin{equation}
  \eta(T) = \exp\left( - \gamma T \right),
  \label{eq_viscosity}
\end{equation}
\begin{equation}
  T = \psi = \frac{ \partial^2 \psi }{ \partial y^2 } = 0
  \quad \text{ at } y = 1,
  \label{eq_BC1}
\end{equation}
\begin{equation}
  \frac{\partial T}{\partial y} =\psi =\frac{\partial^2 \psi}{\partial y^2} = 0
  \quad \text{ at } y = 0 ,
  \label{eq_BC0}
\end{equation}
where
$\psi (x,y)$ is the 
stream function,
$\displaystyle \nabla^2
= \frac{ \partial^2 }{ \partial x^2 } + \frac{ \partial^2 }{ \partial y^2 }$
is the two-dimensional Laplacian,
and
$\displaystyle \frac{ \partial (T, \psi) }{ \partial (x,y) } = \frac{ \partial T }{ \partial x } \frac{ \partial \psi }{ \partial y } - \frac{ \partial T }{ \partial y } \frac{ \partial \psi }{ \partial x }$
is the Jacobian
of temperature and stream function.
The Rayleigh (-Roberts) number is defined as
\begin{equation}
  Ra_0 = \frac{ \alpha \rho g H d^5 }{ \kappa k \eta_0 } ,
  \label{eq_Ra0}
\end{equation}
where
$\alpha$ is the thermal expansion coefficient,
$\rho$ is the density,
$g$ is the gravitational acceleration,
and $k$ is the thermal conductivity. 

\subsection{Numerical method}
\label{subsec_newtonmethod}
Second-order finite differences are used for the numerical formulation.
In particular, the finite difference equation proposed by \cite{Andrews1972}
is used for the Stokes equation \eqref{eq_stokes}.
The domain is discretized into square grids of length $1/32$ or $1/64$ for all numerical calculations.
Most calculations use the resolution of $1/32$. 
For some cases where iterative calculations do not converge, 
the resolution of $1/64$ is used.

We have obtained numerical steady solutions by the Newton method in the same procedure as that of \cite{Okuda2023}.
Let the right-hand-side of the energy equation \eqref{eq_energy} be $F(T)$,
so that a steady solution $T^\prime$ satisfies $F(T^\prime) = 0$.
$F(T)$ is a function of $T(x,y)$
because $\psi(x,y)$ is obtained by the linear equation \eqref{eq_stokes} when $T(x,y)$ is given.
At each iteration, 
a new estimate $T_{\mathrm{new}}$ of the steady solution
is computed
by solving the first-order Taylor expansion of $F(T)$ around the previous value $T_{\mathrm{old}}$ 
at each position $\vec{r}=(x, y)$ in the domain:
\begin{equation}
  0 = F(T_{\mathrm{old}}) (\vec{r})
    + \frac{\mathrm{d}F}{\mathrm{d}T} ( \vec{r},\vec{\xi} )
      \cdot \left( T_{\mathrm{new}} (\vec{\xi}) - T_{\mathrm{old}} (\vec{\xi}) \right) ,
\end{equation}
where the derivative of $F(T)$ by $T$ is calculated by finite numerical differentiation as
\begin{equation}
  \frac{ \mathrm{d} F}{ \mathrm{d} T } (\vec{r},\vec{\xi})
 = \frac{ F(T + \delta T_\xi) (\vec{r}) - F(T) (\vec{r}) }{ \Delta }, 
  \quad
  \delta T_\xi(\vec{r}) = \left\{
                  \begin{aligned}
                    \Delta \quad (\vec{r} =    \vec{\xi}) , \\
                    0      \quad (\vec{r} \neq \vec{\xi}) .
                  \end{aligned}
                 \right.
\end{equation}
Here, $\vec{\xi}$ is a position at which a small perturbation $\Delta$ is added to $T$.
$\Delta$ is given as $10^{-6}$ in this study.
We set the convergence criterion for the calculation of the Newton method
to ensure that
the temperature update for one iteration step becomes less than $10^{-6}$,
where the typical amplitude of $T$ variations is of the order of $\sim 10^{-1}$.

Steady-state calculations are obtained for the ranges of the parameters
of $1 \leq Ra_0 \leq 10^6$ and $0 \leq \gamma \leq 40$. 
Solutions for $20 \leq \gamma \leq 30$ are explored in detail
in order to focus on the solutions around the intermediate (sluggish lid) regime.
Only solutions with horizontal wavelength 2 are considered.

\subsection{Convergence of steady state calculation}
\begin{figure}
  \centering
  \includegraphics[width=0.49\columnwidth]{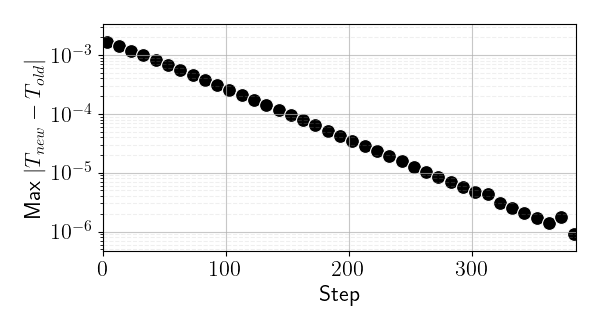}
  \includegraphics[width=0.49\columnwidth]{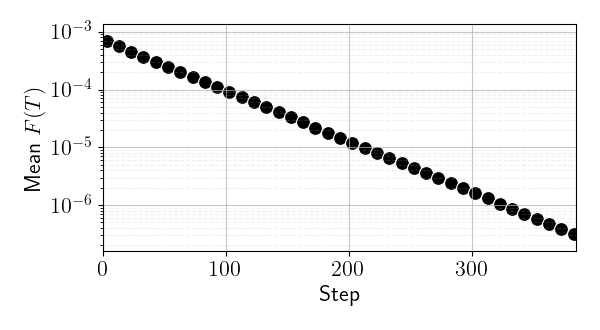}
  \caption{Example of convergence of the Newton method calculation. 
  The maximum norm of temperature update $T_{\mathrm{new}} - T_{\mathrm{old}}$ by one step of iteration (left) and 
  the averaged values of $F(T)$ (the right-hand side of eq.~\ref{eq_energy}) over the whole layer (right) are shown for every ten steps of iteration.
  }
  \label{fig_convergence}
\end{figure}

The convergence of an iterative calculation for a steady solution is shown in
\figref{fig_convergence}.
The calculation runs until the maximum norm of $T_{\mathrm{new}} - T_{\mathrm{old}}$ (left panel) drops below the chosen convergence criterion, \num{e-6}.
The right panel shows the average values of $F(T)$ in the whole domain.
As the calculation converges, 
$F(T)$ decreases by several orders of magnitude.

\subsection{Useful parameters}
\label{subsec_parameters}
At large viscosity contrast, 
convection is in the stagnant lid regime where most of the temperature and viscosity contrasts are concentrated in the lid. 
Since the Rayleigh number $Ra_0$ (eq.~\ref{eq_Ra0}) is constructed using the viscosity at the upper boundary, 
it does not properly characterize the dynamics of the layer where convection actually occurs, below the lid. 
We therefore introduce the bottom Rayleigh number,
$Ra_b$, defined using the viscosity at the bottom boundary:
\begin{equation}
  Ra_b = \frac{ \alpha \rho g H d^5 }{ \kappa k \eta_b }
       = Ra_0 \frac{ \eta_0 }{ \eta_b }
       = Ra_0 \exp( \gamma T_b ),
  \label{eq_Rab}
\end{equation}
with
$\eta_b = \eta_0 \exp( - \gamma T_b )$
and
$T_b$ the averaged temperature at the bottom boundary.
$T_b$ is close to the temperature in the isothermal core
because there is no thermal boundary layer along the adiabatic bottom boundary.

Compared to the situation when a total temperature difference is imposed across the layer, 
the parameter $\gamma$ is not sufficient to quantify the viscosity contrast across the layer, which depends on the output parameter $T_b$.
The viscosity contrast ratio is defined as
\begin{equation}
  r_\eta = \frac{ \eta_0 }{ \eta_b } = \exp(\gamma T_b) .
  \label{eq_reta}
\end{equation}

The Nusselt number is defined using dimensionless variables as
\begin{equation}
  Nu = \frac{ 1 }{ T_b },
  \label{eq_Nu}
\end{equation}
which is the same definition as \cite{SolomatovMoresi2000}.
This is simply the reciprocal of the internal temperature of the convective region.
Note that $Nu=2$ corresponds to the conductive solution without motion. 

\section{Results}
\label{sec_result}

\subsection{Steady state solutions and regime classification}
\label{subsec_steadysolutions}

The obtained steady solutions are classified into three regimes, mobile, sluggish lid and stagnant lid regimes with the top surface mobility $M$ \citep{Tackley2000},
which is defined as the ratio of the root mean square of horizontal velocity at the top surface $u_{\mathrm{surface,rms}}$ to that in the whole layer $u_{\mathrm{rms}}$.  
\begin{equation}
  M = u_{\mathrm{surface,rms}} / u_{\mathrm{rms}} .
\end{equation}
Then we define mobile, sluggish lid, and stagnant lid regimes as
$M > 0.90$, $0.90 \geq M > 0.02$, and $0.02 \geq M$, respectively.
The threshold values are somewhat arbitrary.
The different regimes are expected to follow different scaling relationships between the dimensionless heat flux, the Nusselt number, and the parameters, the Rayleigh number and the temperature-dependence strength of viscosity. Scaling laws for the stagnant lid and mobile regimes are presented in \figref{fig_Nu-Rabdiagram} in Section \ref{subsec_SVCST}, and we adjusted the threshold values of $M$ such that they also match the validity limits of the scaling laws.

\begin{figure}
  \centering
  \includegraphics[width=0.49\columnwidth]{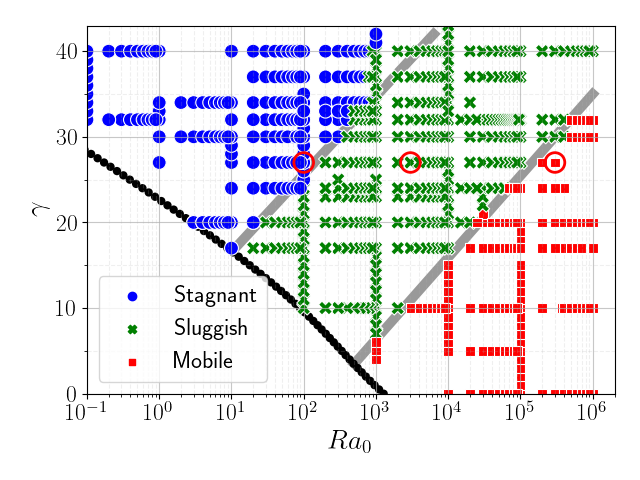}
  \includegraphics[width=0.49\columnwidth]{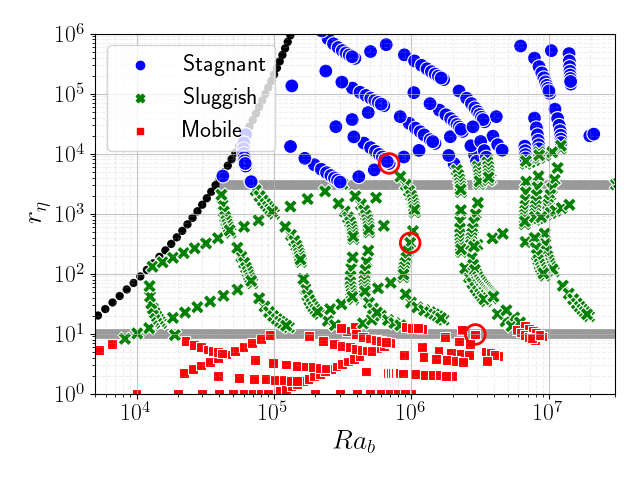}
  \caption{Steady solutions on $Ra_0$--$\gamma$ diagram (left) and $Ra_b$--$r_\eta$ diagram (right). Red, green, and blue symbols correspond to steady solutions classified into mobile regime (Mobile), sluggish lid regime (Sluggish), and stagnant lid regime (Stagnant), respectively. 
  Black circles represent the neutral stability points for the linear stability of the hydrostatic conduction state. 
  Three red circles on each panel identify the cases 
  whose convective structures are shown in Figures \ref{fig_Tempfield} and \ref{fig_verticalprofile}.
  The grey lines indicate the rough regime boundaries.
  } 
  \label{fig_regimediagram}
\end{figure}

All steady-state solutions obtained in this study
are shown as regime diagrams in \figref{fig_regimediagram}.
The left panel shows the diagram of $Ra_0$ and $\gamma$ space, 
both of which being the input control parameters.
The right panel shows $Ra_b$--$r_\eta$ diagram, 
which is useful to compare the result with that of the bottom heating cases.
The black solid circles on both panels are the neutral values of the linear stability analysis for the hydrostatic state.
The neutral stability is obtained following a classical approach \citep[e.g.][]{Chandrasekhar, Stengel1982, Labrosse_etal2018}.
Infinitesimal perturbations of the various variables around their steady conduction solution are written as the product of a function of the vertical coordinate, $F(y)$, a harmonic function of the horizontal coordinate, $e^{ikx}$, and an exponential time function, $\exp(\sigma t)$. The linearized equations for the perturbations are then transformed into an eigenvalue-eigenvector differential equation for $\sigma$ and $F$ that depends on $Ra_0$, $\gamma$, and $k$. Substituting $k=\pi$ (which means a horizontal wavelength of 2) and each value of $\gamma$, this equation is solved using a finite difference method and the critical Rayleigh number is the lowest value of that parameter that gives a non trivial eigenvector to the eigenvalue $\sigma=0$.

In the $Ra_0$--$\gamma$ diagram,
three regimes appear to be clearly separated.
They are in banded forms
in the direction from small $Ra_0$ and $\gamma$ to large $Ra_0$ and $\gamma$.
The mobile regime is obtained for a low $\gamma$ or a high $Ra_0$,
the stagnant lid regime is obtained for a high $\gamma$ or a low $Ra_0$,
and the sluggish lid regime is in between.
The regime boundaries in the $Ra_b$--$r_\eta$ diagram 
correspond to nearly constant values of $r_\eta$, 
$r_\eta \approx 1 \times 10^1$ for the mobile--sluggish lid regimes boundary, and 
$r_\eta$ slightly increasing from $3 \times 10^3$ to $1 \times 10^4$ for $Ra_b$ going from $4 \times 10^4$ to $1 \times 10^7$ for the sluggish lid--stagnant lid regimes boundary.
The value of $r_\eta$ for the sluggish--stagnant boundary is similar to
that obtained by \cite{Huttig2011}
of $r_\eta \approx 10^4$ for internal heating convection in a spherical shell with a radius ratio of 0.55. As shown by \citet{JAVAHERI2024}, this viscosity ratio can strongly depend on the aspect ratio of the spherical shell, at least for the bottom heating situation they considered.

The distribution of the three regimes for internal heating convection shown in the right panel of \figref{fig_regimediagram}
is qualitatively similar to the results for the bottom heating
although the values of the regime boundaries are slightly different. 
\cite{Solomatov1995} and \cite{Kameyama2000}
studied bottom heating convection of 2-D temperature-dependent viscous fluid. 
They showed that the sluggish--stagnant boundary is at $r_\eta=10^4$.
The mobile--sluggish boundary in \cite{Solomatov1995} increases slightly with the Rayleigh number, 
from $r_\eta \approx 10^0$ at $Ra_b =10^3$ to $r_\eta\approx10^3$ at  $Ra_b=10^{11}$, and that in \cite{Kameyama2000} is independent of $Ra_b$ but the value of $r_\eta\approx10^{3.5}$ is different from the one we get, $r_\eta \approx 10^1$. 
\cite{Okuda2023} showed the regime boundaries of steady-state bottom heating convection with a horizontal wavelength of 2
as $r_\eta = e^{1.5} \approx 5 $ for the mobile--sluggish boundary
and $r_\eta = e^{8.0} \approx 3 \times 10^3$ for the sluggish--stagnant boundary.

\begin{figure}
  \centering
  \begin{tabular}{cc}
  Temperature & Stream function \\
  \multicolumn{2}{c}{Mobile regime ($Ra_0 = 3 \times 10^5, \gamma = 27$)} \\
  \includegraphics[width=0.45\columnwidth]{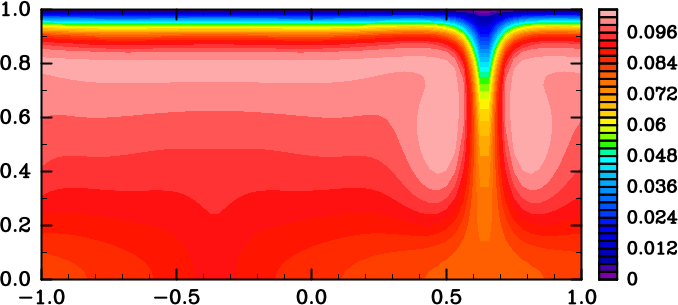} &
  \includegraphics[width=0.45\columnwidth]{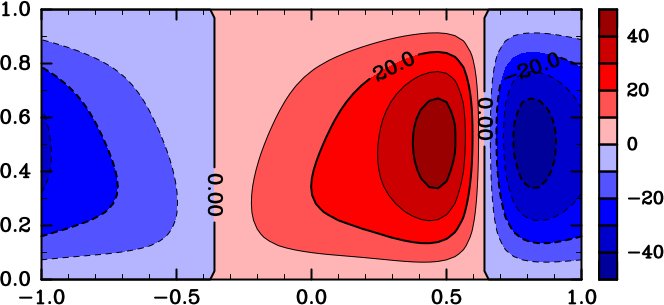} \\
  \\
  \multicolumn{2}{c}{Sluggish lid regime ($Ra_0 = 3 \times 10^3, \gamma = 27$)} \\
  \includegraphics[width=0.45\columnwidth]{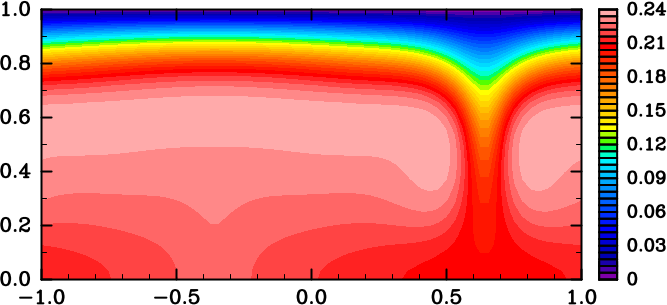} & 
  \includegraphics[width=0.45\columnwidth]{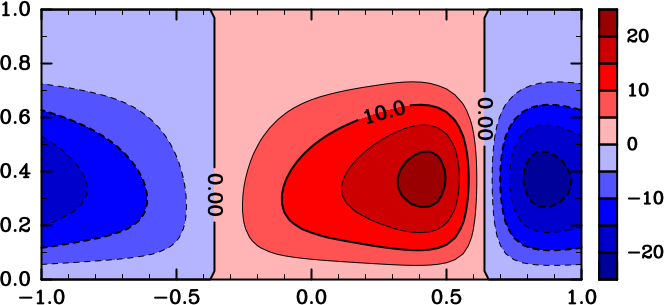} \\
  \\
  \multicolumn{2}{c}{Stagnant lid regime ($Ra_0 = 1 \times 10^2, \gamma = 27$)} \\
  \includegraphics[width=0.45\columnwidth]{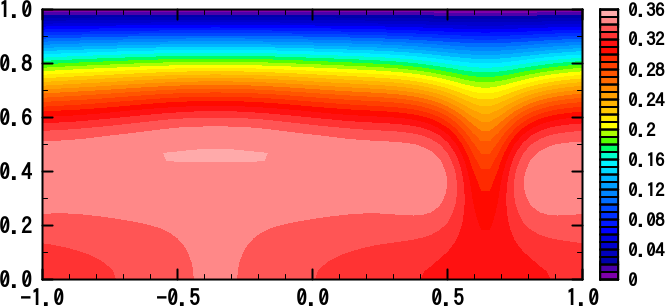} &
  \includegraphics[width=0.45\columnwidth]{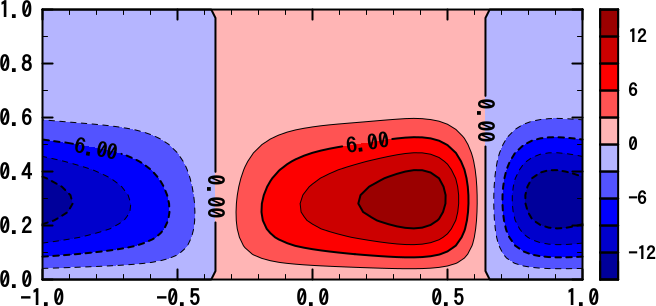} \\
  \end{tabular}
  \caption{Examples of steady solutions of the mobile regime (top), the sluggish lid regime (center), and the stagnant lid regime (bottom). Temperature and stream function are shown in the left and right panels, respectively.
  Parameters $Ra_0$ and $\gamma$ are indicated above each pair of panels.
  }
  \label{fig_Tempfield}
\end{figure}
\begin{figure}
  \centering
    \begin{tabular}{ccc}
     \multicolumn{3}{c}{Mobile regime ($Ra_0 = 3 \times 10^5, \gamma = 27$)} \\
    \includegraphics[height=0.2\textheight]{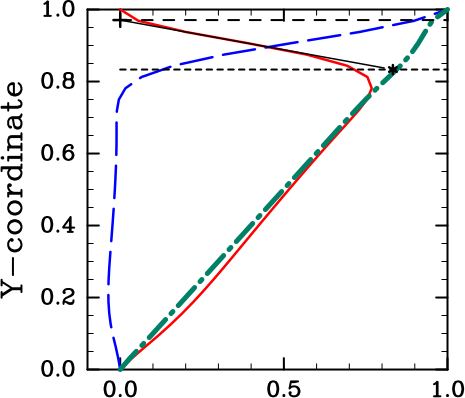}
    &  \includegraphics[height=0.2\textheight]{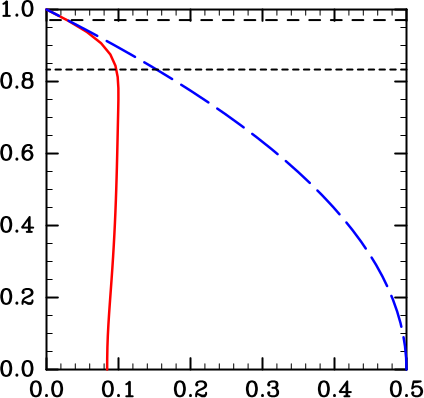}
    &  \includegraphics[height=0.2\textheight]{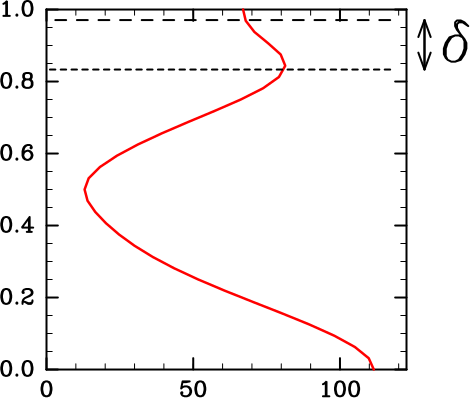}
    \\[0.7em]
    \multicolumn{3}{c}{Sluggish lid regime ($Ra_0 = 3 \times 10^3, \gamma = 27$)} \\
    \includegraphics[height=0.2\textheight]{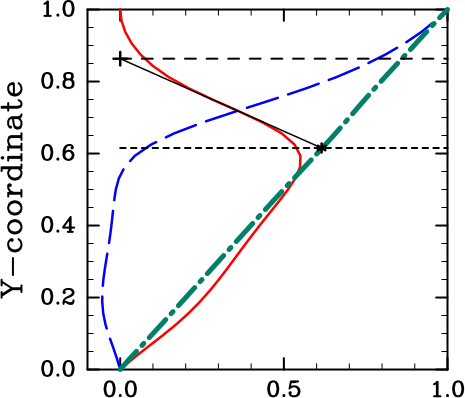}
    &  \includegraphics[height=0.2\textheight]{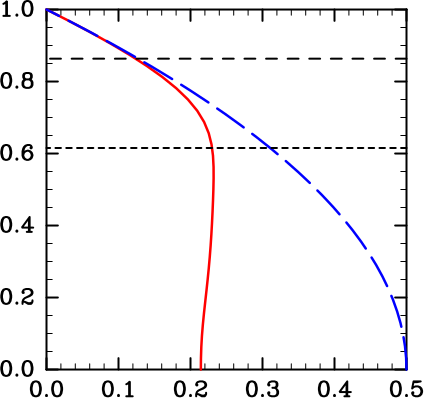}
    &  \includegraphics[height=0.2\textheight]{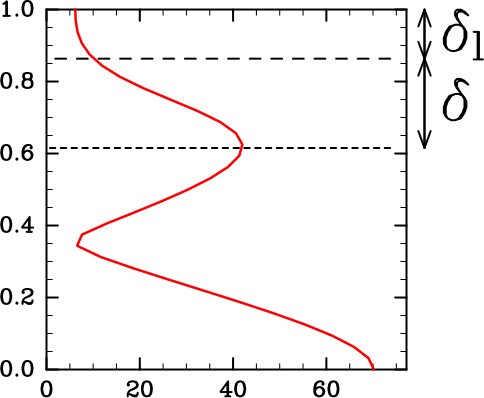}
    \\[0.7em]
    \multicolumn{3}{c}{Stagnant lid regime ($Ra_0 = 1 \times 10^2, \gamma = 27$)} \\
    \includegraphics[height=0.2\textheight]{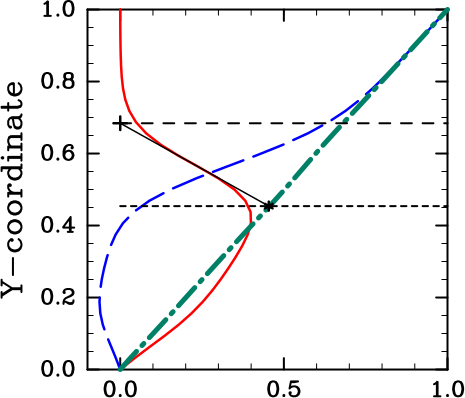}
    &  \includegraphics[height=0.2\textheight]{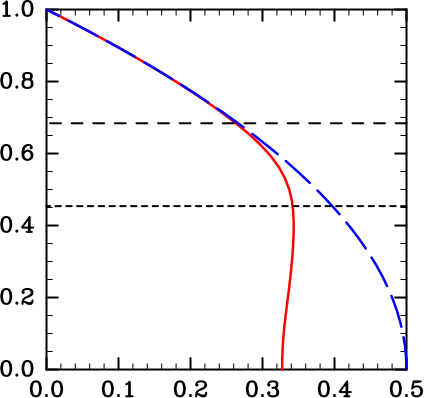}
    &  \includegraphics[height=0.2\textheight]{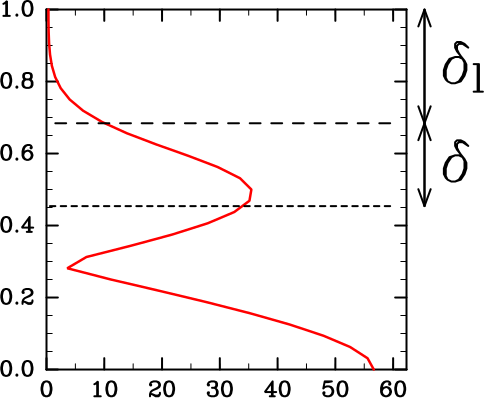}
    \\
    Heat flux & Temperature & $u_{\mathrm{rms}}$
    \end{tabular}
  \caption{
    Vertical profiles of several variables for steady solutions
    of the mobile regime (top), the sluggish lid regime (middle), and the stagnant lid regime (bottom). 
    Left panels show horizontally averaged advection heat flux
    $\overline{vT}$ (red solid lines),  diffusion heat flux $\overline{\partial_y T}$ (blue broken lines), 
    and total heat flux (green dash-dotted lines).
    The center column displays the horizontally averaged temperature $\overline{T}$ (red solid line)
    and the static conductive temperature (blue dashed line). 
    Right panels show the root mean square of the horizontal velocity $u$ for each height, 
    $u_{\mathrm{rms}}$. 
    The horizontal dashed and dotted lines in each panel
    show the boundaries between the conductive lid, active boundary layer
    and convective core.
    They are obtained as crossing between straight-line fits to the advection profile in each region (see text for details).
    The thicknesses $\delta_l$ and $\delta$ of the lid and active boundary layer, respectively, are defined on the right-hand-side of the figure.
    }
  \label{fig_verticalprofile}
\end{figure}

Typical solutions of finite-amplitude convection in each regime
at the points
indicated by red circles in \figref{fig_regimediagram}
are shown in Figures \ref{fig_Tempfield} and \ref{fig_verticalprofile}.
\figref{fig_Tempfield} shows temperature (left) and stream function (right) fields
of the mobile regime (top), the sluggish lid regime (middle), and the stagnant lid regime (bottom).
A cold plume with strong downward flow develops
because the thermal boundary layer exists only along the top surface.
The thickness of the conductive layer above the convective region
increases from the mobile regime solutions to the stagnant lid regime solutions.

The vertical profiles of 
the horizontally averaged 
advection flux $\overline{vT}$ (red solid lines) and
diffusion flux $\overline{\partial_y T}$ (blue broken lines)
are shown in the left panels of \figref{fig_verticalprofile}.
The over-line means horizontal average 
such that $\overline{f}(y)=(1/2)\int_0^2 f(x,y) \mathrm{d}x$. 
The green dash-dotted lines are 
total heat flow across each horizontal plane.
By virtue of the conservation of energy (eq.~\ref{eq_energy}) in steady state, it is
$\overline{vT} + \overline{\partial_y T} = y$, 
which is equal to the total heat production below height $y$.
Two horizontal thin dashed lines for each left panel divide
from top to bottom
a conductive lid (thickness $\delta_l$),
an active boundary layer (thickness $\delta$),
and an isothermal core.
To objectively determine the height of these boundaries, 
the advection heat flux
is approximated as a piecewise affine profile.
Let $\overline{vT}$ in the active boundary layer $\overline{vT}=ay+b$
be defined by the tangent line at the
inflection point
(thin black solid lines).
The stagnant lid being defined as a region where $\overline{vT} = 0$, 
its bottom boundary
is the contact point of the advection heat flux on the $y$ axis
($ay+b=0$, shown by $+$ in the left panels).
The bottom of the active boundary layer is defined as the
intersection of the advection tangent
with the total heat flux, 
so the height is where $ay+b=y$ is satisfied ($*$ in the left panels).
The conductive lid, 
boundary layer, and
isothermal core
defined in this way
can be considered as regions where $\overline{vT} = 0$,
$\overline{vT}$ and $\overline{\partial_y T}$ are comparable,
and $\overline{\partial_y T} = 0$,
respectively.

The center and right panels of \figref{fig_verticalprofile} show
the horizontal mean temperature $\overline{T}$ (red solid lines in the center columns),
the conductive temperature (blue dashed lines, equal to $(1-y^2)/2$), 
and the root mean square of horizontal velocity at each height $u_{\mathrm{rms}}$ (red lines in the right columns).
From the profiles of $u_{\mathrm{rms}}$,
we can find that
the velocity gradient in the conductive lid is small
regardless of whether the lid moves or not. 
The velocity gradient is large in the boundary layer and in the isothermal core, 
but the temperature gradient driving the convective flow is concentrated in the boundary layer. 

\subsection{Scaling functions of the mobile regime and the stagnant lid regime}
\label{subsec_SVCST}

Here, we examine whether the steady solutions obtained in the mobile regime and the stagnant lid regime
are consistent with the scaling relations proposed by previous studies. 
\cite{Grasset1998} (hereafter GP98)
analyzed the data obtained by their numerical time integrations
and obtained the following relations, with replacement of the notation, 
for the solutions in the no-lid regime (eq.~14 in GP98) 
and in the conductive lid regime (eq.~29 in GP98), respectively.
\begin{align}
  T_b &= a_m r_\eta^{\alpha_m} Ra_b^{-\beta_m},
  \label{eq_nolid}
  \\
  \tilde{T}_b
  &= a_s \widetilde{Ra}_b^{-\beta_s}.
  \label{eq_condlid}
\end{align}
Here $\alpha_m$, $\beta_m$, and $\beta_s$ are the exponents that are determined later by fitting these functions to our numerical steady solutions.
The tilde expresses the rescaled variable for the stagnant lid regime
using a length scale equal to the height of the convective region $1-\delta_l$, 
expecting the same scaling relation inside the convective region as for iso-viscous convection,
where $\delta_l$ is the thickness of the conductive lid defined by advection flux profile (see \figref{fig_verticalprofile}).
Specifically,
assuming that the viscosity ratio across the active boundary layer, $r_{\eta \mathrm{eff}}$, is fixed according to GP98, 
and applying the energy balance in the conductive lid, 
it is written as
$ 1-\delta_l = \sqrt{ 1 - 2 (T_b - \ln(r_{\eta \mathrm{eff}})/\gamma) } $
(eqs.~23 and 28 of GP98) and 
\begin{equation}
    \tilde{T}_b    = \frac{ \ln(r_{\eta \mathrm{eff}}) }{ \gamma (1-\delta_l)^2 } , \quad 
    \widetilde{Ra}_b     = Ra_b (1-\delta_l)^5 , \quad 
    \tilde{\delta} = \frac{\delta}{ 1-\delta_l } , \quad 
    \tilde{u}      = u (1-\delta_l) .
    \label{eq_STvariables}
\end{equation}
Using the scaling functions \eqref{eq_nolid} and \eqref{eq_condlid},
$T_b$
is related to the input parameter $Ra_0$ and $\gamma$ with the following equations, respectively for the mobile and stagnant lid regimes:
\begin{align}
    T_b \exp\left( - \alpha_m \gamma T_b \right)
    &= a_m Ra_0^{-\beta_m} \exp( -\beta_m \gamma T_b ) , 
    \label{eq_Ra0function_SVC} \\
    \frac{ \ln{(r_{\eta \mathrm{eff}})} }{ \gamma }
    \left[ 1 - 2 \left( T_b - \frac{\ln{(r_{\eta \mathrm{eff}})}}{\gamma} \right) \right]^{\frac{5}{2}\beta_s -1} 
    &= a_s Ra_0^{-\beta_s} \exp \left( -\beta_s \gamma T_b \right) .
    \label{eq_Ra0function_ST}
\end{align}
Alternatively, the right-hand-side of both equations can be written as 
$a Ra_b^{-\beta}$, which means that the scaling laws can be written with either $Ra_0$ or $Ra_b$.
\begin{figure}
  \centering
  \includegraphics[width=0.7\columnwidth]{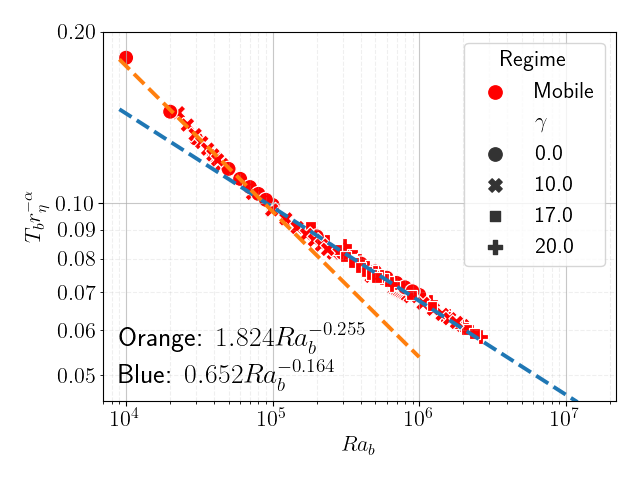}
  \caption{
    $T_b r^{-\alpha_m}$--$Ra_b^{-\beta}$ diagram for the solutions in the mobile regime.
    Two dash-dotted lines are the functions of eq.~\eqref{eq_nolid}
    fitted to solutions of $Ra_b < 10^5$ (orange) and $Ra_b > 10^5$ (blue). 
    The best fit values of $\alpha_m$, $\beta_m$, and $a_m$ in eq.~\eqref{eq_nolid} are obtained as 
    $\alpha_m=0.167, \beta_m=0.255$, and $a_m=1.82$ for $Ra_b < 10^5$ and 
    $\alpha_m=0.159, \beta_m=0.164$, and $a_m=0.652$ for $Ra_b \geq 10^5$. 
    The values $T_b r_\eta^{-\alpha_m}$ for each solution use the corresponding value of $\alpha_m$ according to the value of $Ra_b$.}
  \label{fig_SVCfuncGP98}
\end{figure}
%

\figref{fig_SVCfuncGP98} shows 
solutions in the $T_b r_\eta^{-\alpha_m}$--$Ra_b$ space for the mobile regime
to examine whether our results are consistent 
with eq.~\eqref{eq_nolid} or not. 
We fitted the function \eqref{eq_nolid} to the mobile regime solutions 
to obtain the best fit values
of $\alpha_m$, $\beta_m$, and $a_m$.
Two different sets of coefficients are derived:
$\alpha_m=0.167$, $\beta_m=0.255$, and $a_m=1.82$ for $Ra_b < 10^5$,
and 
$\alpha_m=0.159$, $\beta_m=0.164$, and $a_m=0.652$ for $Ra_b \geq 10^5$. 
Two dashed lines correspond to function \eqref{eq_nolid} with these sets of coefficients.

\begin{figure}[htbp]
  \centering
  \includegraphics[width=0.7\columnwidth]{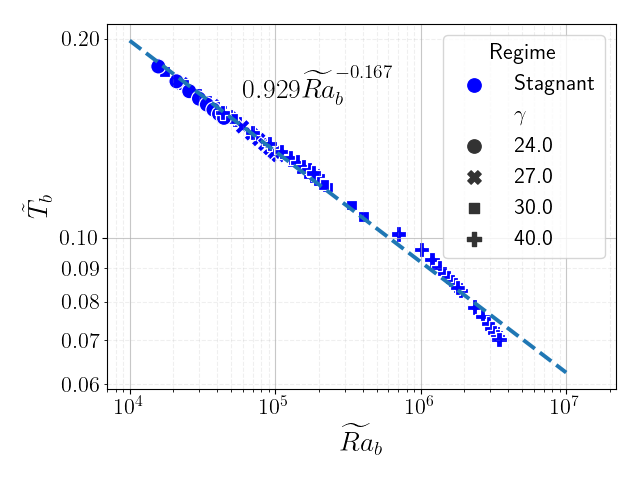}
  \caption{
    $\tilde{T}_b$-$\widetilde{Ra}_b$ plot for the solutions in stagnant lid regime.
    The dash-dotted line is the result of the power-law fitting 
    of the function \eqref{eq_condlid} to data of stagnant lid regime solutions with different values of $Ra_0$ and some fixed values of $\gamma$. 
    By setting $r_{\eta \mathrm{eff}}=5.4$ in eq.~\eqref{eq_STvariables} based on our observation of viscosity contrast below the stagnant lid (\figref{fig_retaeff-Rab}), 
    the obtained best fit values are $\beta_s=0.167$ and $a_s=0.929$ in eq.~\eqref{eq_condlid}. 
    }
  \label{fig_STfuncGP98}
\end{figure}

\figref{fig_STfuncGP98} shows $\tilde{T}_b$-$\widetilde{Ra}_b$ plot for the solutions in the stagnant lid regime. 
$r_{\eta \mathrm{eff}}$ in eq.~\eqref{eq_STvariables} is fixed to $5.4$
based on the observation of our stagnant lid regime solutions, which will be described in \figref{fig_retaeff-Rab} in Section \ref{subsec_characterizatoinTR}.
We performed the power law fitting 
to obtain the best fit value of $\beta_s$.
The resulting dashed line 
in \figref{fig_STfuncGP98} indicates that $\beta_s=0.167$, 
which is very similar to the mobile regime solutions of 
$Ra_b \gtrsim 10^5$. 
The power-law fitting explains well the solutions with $\widetilde{Ra}_b \lesssim 10^6$. 
On the other hand, there appears to be a misfit for the solutions with $\widetilde{Ra}_b \gtrsim 10^6$ ($\gamma=40$), possibly due to the loose regime classification with mobility $M$. 
These solutions are close to the boundary with the sluggish lid regime and could be classified in that regime
if a different threshold value of $M$ were chosen
or a different criterion for lid mobility were used.

The mobile regime solutions for $Ra_b \lesssim 10^5$ 
follow the power law with $\beta_m = 0.255$, 
which is consistent with the $Ra_b^{-1/4}$ power law
obtained by the previous studies on iso-viscous convection
\citep[e.g.][]{Parmentier1994, Geodynamics2014}
and the no-lid regime of GP98.
On the other hand,
solutions of $Ra_b \gtrsim 10^5$ 
follow the power law functions of 
$\beta \approx 1/6$
for both the mobile and stagnant lid regimes.
The difference between time-dependent
and steady solutions may cause this smaller exponent of power law.
Section \ref{subsec_derivationofscaling} discusses
this issue in detail. 

\begin{figure}
  \centering
  \includegraphics[width=0.8\columnwidth]{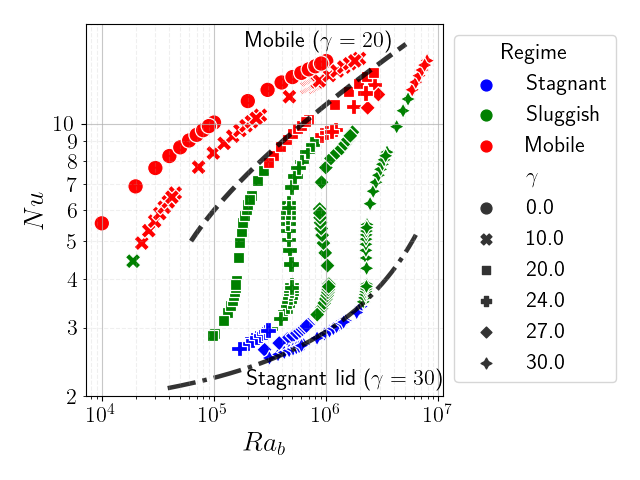}
  \caption{Steady solutions for several values of $\gamma$ on $Nu$--$Ra_b$ plane.
  Colors represent the regime of each solution. 
  The Nusselt number $Nu$ and the Rayleigh number $Ra_b$ are defined by eqs.~\eqref{eq_Nu} and \eqref{eq_Rab}.
  The scaling relations of the stagnant lid regime (dash-dotted, eq.~\ref{eq_Ra0function_ST}) 
  and the mobile regime (dashed, eq.~\ref{eq_Ra0function_SVC}) 
  are overlaid. The coefficient and exponents in these equations are determined
  in Section \ref{subsec_SVCST}, and $\gamma$ is fixed as indicated values near each curve.
  Note that $Ra_b$ does not necessarily increase monotonically even if $Ra_0$ increases
  because $Ra_b$ depends on the resulting variable $T_b$.
  }
  \label{fig_Nu-Rabdiagram}
\end{figure}

Using the coefficients $\alpha_m=0.159$ and $\beta_m=0.164$ for the mobile regime, and $\beta_s=0.167$ for the stagnant lid regime, 
\figref{fig_Nu-Rabdiagram} shows
the scaling relation curves
for $\gamma=20$ of the mobile regime (eq.~\ref{eq_Ra0function_SVC}) and $\gamma=30$ of the stagnant lid regime (eq.~\ref{eq_Ra0function_ST}) in the $Nu$--$Ra_b$ space together with steady solutions for several values of $\gamma$ as examples. 
$Nu$ and $Ra_b$ are given by eqs.~\eqref{eq_Nu} and \eqref{eq_Rab}.
The functions of the mobile and stagnant lid regimes fit well to the solutions of respective regimes, including other setting of $\gamma$.
Note that
solutions in the sluggish lid regime do not follow any of the scaling functions obtained for the mobile and stagnant lid regimes,
for any fixed value of $\gamma$, 
even though the curves for other values of $\gamma$ are not displayed in \figref{fig_Nu-Rabdiagram}. 
The data of sluggish lid regime solutions show that
$Nu$ increases strongly with a small change of $Ra_b$
for each value of $\gamma$.
To explore each $\gamma$ branch, we vary monotonically the input control parameter $Ra_0$ but $Ra_b$, which depends on the value of $T_b$ actually realized, does not increase monotonically.
Solutions in the sluggish lid regime are characterized in detail
in Section \ref{subsec_characterizatoinTR}.

\subsection{Characterization of the sluggish lid regime}
\label{subsec_characterizatoinTR}

\begin{figure}
  \centering
  \includegraphics[width=0.7\columnwidth]{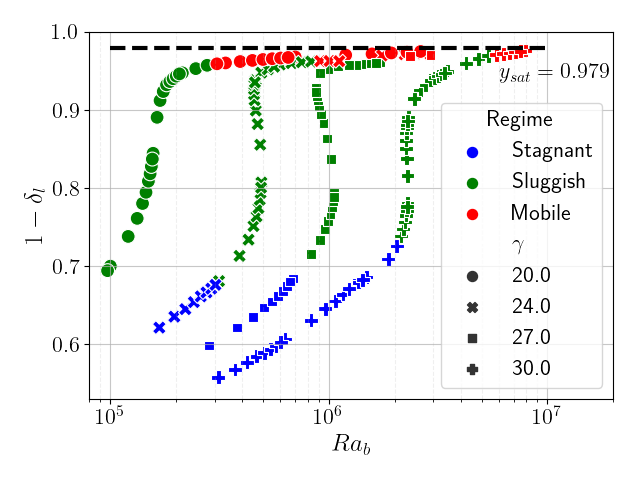}
  \caption{
  The height of the convective region 
  $1-\delta_l$ as a function of
  $Ra_b$ for the solutions of $\gamma=20, 24, 27$, and $30$. 
  $\delta_l$ is the thickness of the conductive lid defined by advection flux profile (see \figref{fig_verticalprofile}).
  The dashed line shows 
  the maximum value of $1-\delta_l$. }
  \label{fig_y2-Rab}
\end{figure}
\begin{figure}
  \centering
  \includegraphics[width=0.7\columnwidth]{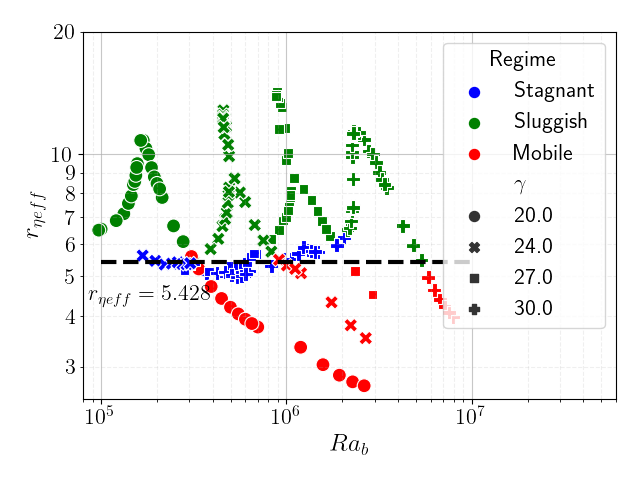}
  \caption{
  The viscosity contrast ratio 
  across the boundary layer between the isothermal core and
  the bottom of the conductive lid, 
  $r_{\eta \mathrm{eff}}$, 
  as a function of $Ra_b$ for the solutions of 
  $\gamma=20, 24, 27,$ and $30$.
  The dashed line expresses the average value of $r_{\eta \mathrm{eff}}$ for the stagnant lid regime solutions. }
  \label{fig_retaeff-Rab}
\end{figure}

To specify different characteristics of the sluggish lid regime,
two variables that are related to thermal structures of convection
are shown in Figures \ref{fig_y2-Rab} and \ref{fig_retaeff-Rab}.

\figref{fig_y2-Rab} shows the height of the convective region, $1-\delta_l$,
as a function of $Ra_b$
for several values of $\gamma$,
where $\delta_l$ is the thickness of the conductive lid defined
in Section \ref{subsec_steadysolutions}
(see \figref{fig_verticalprofile}).
For any given value of $\gamma$, 
starting in the stagnant lid regime at low values of $Ra_b$, 
the height of the convective region increases steadily when increasing $Ra_0$ and therefore $Ra_b$. 
This corresponds to an increase of convection vigor that tends to gain over the stagnant lid. 
When entering the sluggish lid regime, 
increasing $Ra_0$ leads to a strong increase of $1-\delta_l$ (decrease of $\delta_l$ toward 0) at a nearly constant value of $Ra_b$.
Upon further increase of the Rayleigh number, 
the system enters in the mobile regime 
where the thickness of the convectively active region gets close to unity.
However, $1-\delta_l$ 
does not reach $1$ and saturate at $y_{sat}$, contrary to the physical insight that the conducting lid should disappear in the mobile regime.
This is caused by our definition of the conductive lid
by approximating $\overline{vT}$  with the piecewise linear profile,
while $\overline{vT}$ actually behaves as a quadratic function of the distance from the top surface near $y=1$. 

\figref{fig_retaeff-Rab} shows
the viscosity contrast ratio
across the boundary layer
between the active isothermal core
and the bottom of the conductive lid, 
$r_{\eta \mathrm{eff}}$, as a function of $Ra_b$
for the same solutions as used in
\figref{fig_y2-Rab}.
The most remarkable feature is the high values
of $r_{\eta \mathrm{eff}}$ for the solutions in the sluggish lid regime
compared to those for the solutions in the mobile and stagnant lid regimes.
We observe that the viscosity contrast
is almost constant, $r_{\eta \mathrm{eff}} \approx 5.4$,
for the solutions in the stagnant lid regime,
which is consistent with the results of the previous studies.  
\cite{Davaille1993}
found a constant viscosity contrast ratio 
across the convective region below the conductive lid. 
GP98 analyzed the solutions of the stagnant lid regime
and obtained the best-fit value of $e^{2.23} \approx 9.3$, 
which is about twice our result.
When the solutions change from the stagnant lid regime to the sluggish lid regime, 
the values of $r_{\eta \mathrm{eff}}$ increase,
and reach a maximum in the middle of the sluggish lid regime for each value of $\gamma$.
Upon further increase of $Ra_b$, $r_{\eta \mathrm{eff}}$ decreases
and the system leaves the sluggish lid regime for the mobile regime,
where $r_{\eta \mathrm{eff}}$ becomes smaller than the one in the stagnant lid regime. 

From these results,
convection in the sluggish lid regime can be characterized
as a rapid increase of $Nu$ with $Ra_b$ 
compared to the scaling relations of the mobile and stagnant lid regimes,
where the height of convection increases
and $Ra_b$ remains almost constant even if $Ra_0$ is changed.
In the sluggish lid regime,
a larger value of $r_{\eta \mathrm{eff}}$ is reached compared to the nearly constant value obtained in the stagnant lid regime.
The assumption of constant $r_{\eta \mathrm{eff}}$ for the stagnant lid regime by GP98 is based on the existence of a motionless conductive lid, 
so that the feature of larger $r_{\eta \mathrm{eff}}$ for the sluggish lid regime may relate to the mobility of a conductive lid, 
which will be discussed in Section \ref{subsec_rigid}. 

\section{Discussions}
\label{sec_discussion}

\subsection{Scaling relations for steady convection with a strong downward plume}
\label{subsec_derivationofscaling}

\begin{table}
    \centering
    \begin{tabular}{cccc}
              & Numerical results & Theory & GP98 \\ \hline
        \multicolumn{4}{l}{Mobile regime ($Ra_b<10^5$)} \\
        $\delta, T_b$ & $r_\eta^{0.167} Ra_b^{-0.255}$ & 
        & 
        $r_\eta^{0.155} Ra_b^{-0.227} $ \\
        \hline
        \multicolumn{4}{l}{Mobile regime ($Ra_b > 10^5$)} \\
        $\delta, T_b$ & $r_\eta^{0.159} Ra_b^{-0.164}$ &
        $r_\eta^{1/6} Ra_b^{-1/6}$ & 
        $r_\eta^{0.155} Ra_b^{-0.227} $\\
        $u$   & $r_\eta^{-0.30} Ra_b^{0.339}$ & 
        $r_\eta^{-1/3} Ra_b^{1/3}$ & \\
        $\int \Phi \mathrm{d}V$ $\sim \left( u/\delta \right)^2$ & $r_\eta^{-0.92} Ra_b^{1.01} $ &$r_\eta^{-1} Ra_b^1$ 
        &
        \\ \hline\hline
        \multicolumn{4}{l}{Stagnant lid regime} \\
        $\tilde{\delta}, \tilde{T}_b$ & 
        $\widetilde{Ra}_b^{-0.167}$ & $\widetilde{Ra}_b^{-1/6}$ & $Ra_1^{-0.227} $ \\
        $\tilde{u}$                       & 
        $\widetilde{Ra}_b^{ 0.371}$ & $\widetilde{Ra}_b^{ 1/3}$ &                  \\ 
        $\int \widetilde{\Phi} \mathrm{d}\widetilde{V}$ $\sim \left( \tilde{u} / \tilde{\delta} \right)^2$ &
        $\widetilde{Ra}_b^{1.07}$ & $\widetilde{Ra}_b^1$ & \\ \hline
    \end{tabular}
    \caption{Summary of scaling functions for the mobile and stagnant lid regimes.
       $\delta$ is the active boundary layer thickness
       (see \figref{fig_verticalprofile}).
       $u$ is RMS of the horizontal velocity taken at the height where the maximum of $u$ locates.
       Total viscous dissipation $\int \Phi \mathrm{d}V$ scales $\sim (u/\delta)^2$. 
    Variables with a tilde are rescaled for the stagnant lid regime defined by eq.~\eqref{eq_STvariables} using a length of the height of the convective region $1-\delta_l$.
    The functions of the ``Numerical results'' column are derived from 
    Figures \ref{fig_SVCfuncGP98}, \ref{fig_STfuncGP98}, and \ref{fig_u-ra}, 
    while those of the ``Theory'' column are derived in Section \ref{subsec_derivationofscaling}.
    }
    \label{tab_scaling}
\end{table}

Table \ref{tab_scaling} summarizes the fitted power law functions obtained in Section \ref{subsec_SVCST} in
the ``$\delta, T_b$'' and ``$\tilde{\delta}, \tilde{T}_b$'' rows of 
the ``Numerical results'' column.
The functions of the bottom temperature $T_b$ and the rescaled one $\tilde{T}_b$ are referred to Figures \ref{fig_SVCfuncGP98} and \ref{fig_STfuncGP98}, respectively.
In Section \ref{subsec_SVCST},
we found that the power of the Rayleigh number, $\beta_m$ and $\beta_s$, in eqs.~\eqref{eq_nolid} and \eqref{eq_condlid} 
for the high-$Ra_b$ mobile regime and the stagnant lid regime
are different from those proposed by GP98 (``GP98'' column). 
The exponent $\beta_m = 0.255$ of our low-$Ra_b$ mobile regime is almost the same as $0.227$ of GP98, 
while $\beta_m$ and $\beta_s$ are about $1/6$ for other cases of our calculations. 
The power law with $\beta=1/4$ can be explained by the classical boundary layer theory
\citep{Geodynamics2014}, 
while $\beta=1/6$ has not been obtained in previous studies.
We propose the following scaling discussions for the lower exponent of internal heating convection.

\begin{figure}[htbp]
  \centering
  \begin{tabular}{cc}
  Symmetric cell type & Strong downward plume type \\
  ($Ra_0 = 1 \times 10^4, \gamma = 0$) & ($Ra_0 = 1\times 10^6, \gamma = 0$) \\[0.5ex]
  \multicolumn{2}{c}{Temperature} \\
  \includegraphics[width=0.45\columnwidth]{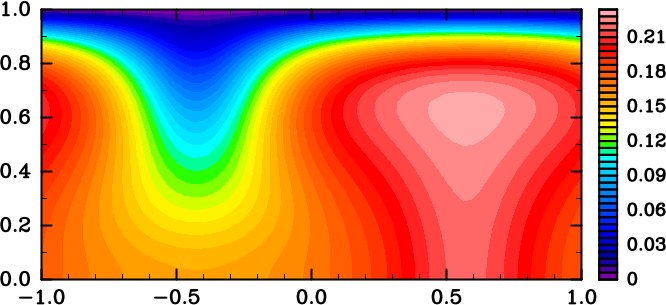} &
  \includegraphics[width=0.45\columnwidth]{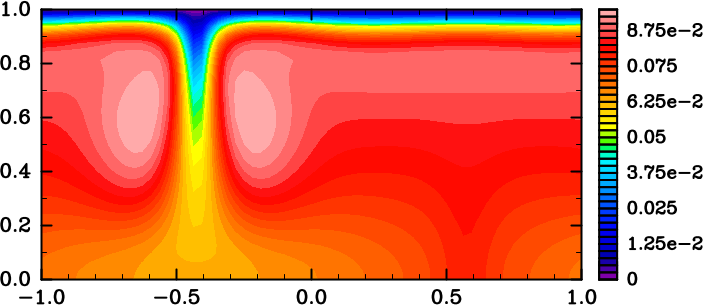} \\
  \\
  \multicolumn{2}{c}{Stream function} \\
  \includegraphics[width=0.45\columnwidth]{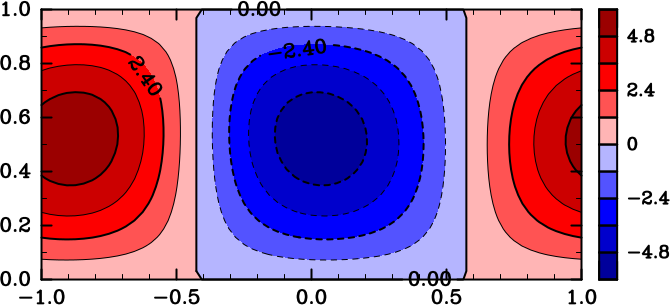} &
  \includegraphics[width=0.45\columnwidth]{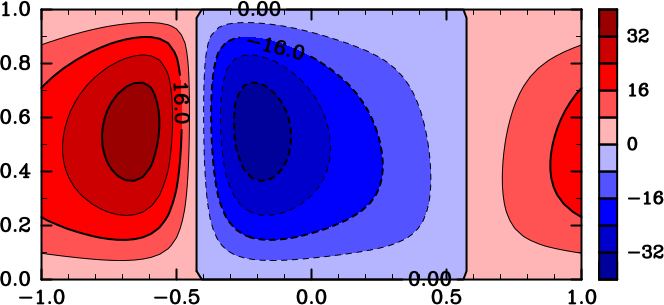} \\
  \\
  \multicolumn{2}{c}{Viscous dissipation} \\
  \includegraphics[width=0.45\columnwidth]{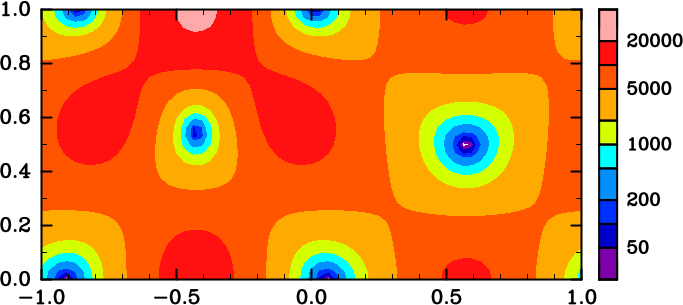} &
  \includegraphics[width=0.45\columnwidth]{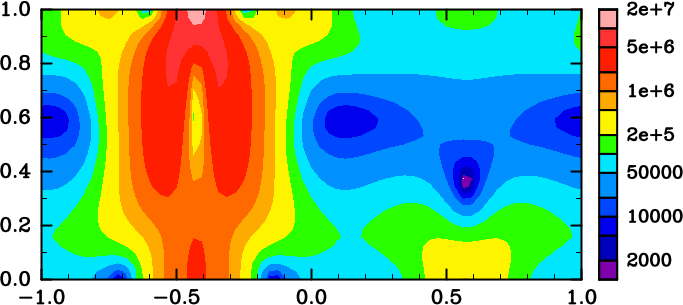}
  \end{tabular}
  \caption{Examples of solutions with two types of convection; 
  symmetric cell convection (left panels, the solution of $\gamma=0$ and $Ra_b = 1.0 \times 10^4$), and 
  convection which has strong downward plume (right panels, $\gamma=0$ and $Ra_b = 1.0 \times 10^6$). 
  The temperature, stream function, and viscous dissipation are shown from top to bottom.
  }
  \label{fig_viscdiss}
\end{figure}

\figref{fig_viscdiss} compares the flow structures and viscous dissipation distribution
for small and large values of the Rayleigh number, $Ra_b=1 \times 10^4$ and $Ra_b=1 \times 10^6$, in the mobile regime. 
Distribution of viscous dissipation is one of the key points for obtaining scaling relations.
Viscous dissipation $\Phi (x,y)$ is defined as
\begin{equation}
    \Phi (x,y) = \sum_{i,j=x,y} \eta 
        \left( \frac{\partial u_i}{\partial x_j} + \frac{\partial u_j}{\partial x_i} \right)
        \frac{\partial u_i}{\partial x_j}.
    \label{eq_viscdiss}
\end{equation}
Steady convection with high $Ra_b \gtrsim 10^5$
typically has a strong and thin downward plume
(right panels of \figref{fig_viscdiss}),
while convection with $Ra_b \lesssim 10^5$ has almost symmetric stream lines (left panel of \figref{fig_viscdiss}). 
Such a symmetric shape of the convection cell has been assumed in the theory leading to the $1/4$ power law for iso-viscous internal heating convection
\citep{Geodynamics2014}.
The difference in the viscous dissipation distributions is clearly seen in the bottom panels of \figref{fig_viscdiss}. 
Similar asymmetric flow was observed by \cite{Mckenzie1974}
with an internal heating configuration, 
and they found a lower exponent than $1/4$ as well.
\cite{SolomatovMoresi2000}
investigated stagnant lid convection by time integration, 
however, 
they did not find a solution with strong asymmetric flow,
and suggested that such solutions are unlikely to be stable. 
The stability of steady solutions in time-dependent calculations will be discussed in Section \ref{subsec_stability}.
Based on these flow and viscous dissipation distributions, 
we will follow and revise the derivation of scaling relations of the classical boundary layer theory
\citep{Geodynamics2014}. 
In order to include the effect of the thinner plumes, the typical horizontal scale of the lateral velocity variations is considered to be similar to the boundary layer thickness rather than the entire wavelength of the convection cells.

First, in the mobile regime, in the upper thermal boundary layer, the balance between horizontal thermal advection, 
$ u\cdot \nabla T \sim u T / \lambda $,
and vertical diffusion, $ \nabla^2 T \sim  T / \delta^2 $
is assumed.
This gives $u \sim \delta^{-2}$, where $u$ and $\delta$ are the typical amplitude of the horizontal velocity and the thickness of the boundary layer below the top surface or conductive lid.
Taking into account the thin downward plume, 
we assume that 
the lateral variations of velocity concentrate on a width proportional to $\delta$
rather than on a width proportional to the wavelength of the convective cells.
The amplitude of the velocity gradient
is evaluated as $u/\delta$, and the total viscous dissipation $\int \Phi \mathrm{d} V \sim \eta (u/\delta)^2\sim \delta^{-6}$. 
Finally, the energy balance relation $\int \Phi \mathrm{d} V = Ra_0\int vT \mathrm{d} V = r_\eta^{-1} Ra_b$ exactly holds, and
$\delta \sim r_\eta^{1/6} Ra_b^{-1/6}$ is obtained.
$T_b$ follows the same scaling as $\delta$
($\delta, T_b$ of the ``Theory'' column for the mobile regime ($Ra_b>10^5$)  in Table \ref{tab_scaling}).
We can follow the same procedure for the stagnant lid regime using the variables rescaled by the height of the convective core. 
It can be seen that the lower exponent of the power of the Rayleigh number
for $T_b$ and $\delta$, $-1/6$,  is successfully derived
($\tilde{\delta}, \tilde{T}_b$ of the ``Theory'' column for the stagnant lid regime in Table \ref{tab_scaling}).
The theoretical scaling relations for other variables are also described in the ``Theory'' column of Table 1. 

\begin{figure}
  \centering
  \includegraphics[width=0.49\columnwidth]{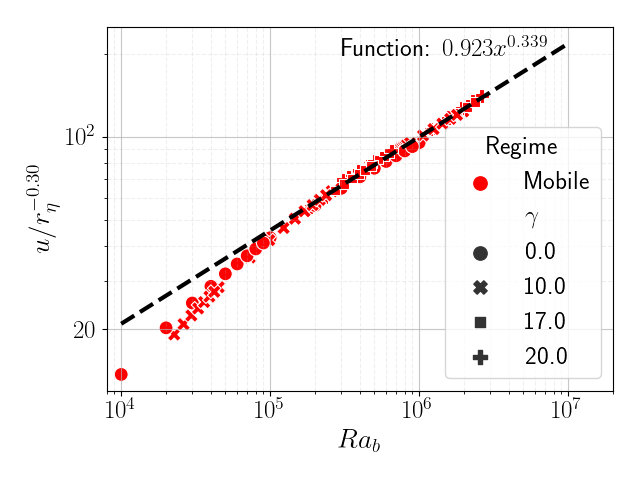}
  \includegraphics[width=0.49\columnwidth]{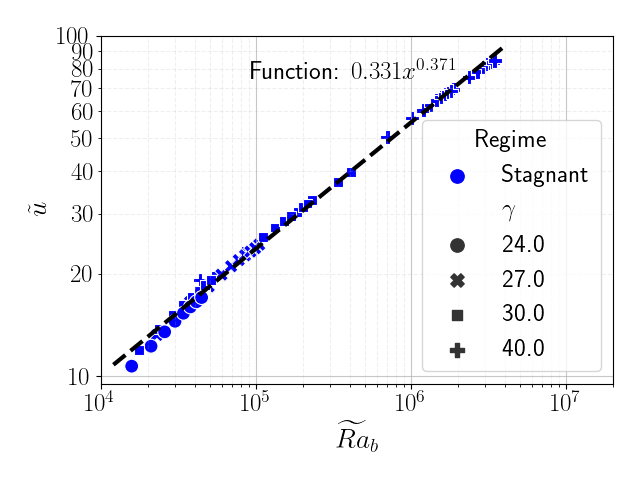}
  \caption{(Left) 
  $u / r_{\eta}^{-0.3}$ versus $Ra_b$ diagram
  for the solutions in the mobile regime, 
  where $u$ is 
  the root mean square of the horizontal velocity 
  taken at the height where the maximum of $u$ is located. 
  The dashed line shows a power law function fitted to the solutions for $Ra_b > 10^5$. 
  (Right) $\tilde{u}$ versus $\widetilde{Ra}_b$ diagram
  for the solutions in the stagnant lid regime. 
  The tilde indicates rescaled variables \eqref{eq_STvariables} in the active convective region.}
  \label{fig_u-ra}
\end{figure}

The scaling relations of the horizontal velocity and total viscous dissipation estimated from our numerical steady solutions are additionally summarized in the column ``Numerical results'' of Table \ref{tab_scaling}.
The typical value of the horizontal velocity, $u$, is calculated as the root mean square of the horizontal velocity at the height where it is largest.
The rescaled variables for the stagnant lid regime (denoted with tilde) are given in eq.~\eqref{eq_STvariables}.
The total viscous dissipation is the $\Phi(x,y)$ integrated in the whole domain for solutions in the mobile regime and in the convective domain (the isothermal core and the boundary layer) for solutions in the stagnant lid regime.
Comparing the columns ``Numerical results'' and ``Theory'' of Table \ref{tab_scaling}, we can see that the theoretical scaling relations of these variables are consistent with those obtained from the steady numerical solutions, although there is a slightly larger discrepancy in the exponent of $\tilde{u}$ of the stagnant lid regime compared to other variables.

To confirm the scaling relations of the horizontal velocity, \figref{fig_u-ra} shows the relations of the horizontal velocity $u$ to the Rayleigh number $Ra_b$ and $\tilde{u}$ to $\widetilde{Ra}_b$.
We can see that the solutions with $Ra_b > 10^5$
lie on the fitted lines and the exponents of the Rayleigh numbers are consistent with those of the theoretical estimates.
Note that the scaling laws are developed for large values of the Rayleigh number, since they require the flow to be characterized by a thin plume. This explains the discrepancy observed between the scaling laws and the numerical results at low values of $Ra_b$.

\subsection{Stability of steady solutions}
\label{subsec_stability}

High $Ra_b$ convection observed in laboratory experiments \citep{Davaille1993}
and time-dependent simulations
\citep[e.g. GP98;][]{SolomatovMoresi2000}
is intermittent with chaotic plumes.
It seems that the width of convection cells
for time-dependent convection
is narrower than that of high $Ra_b$ steady solutions presented here
(e.g.~Figures \ref{fig_Tempfield} or \ref{fig_viscdiss}).
Moreover,
the high values of $r_{\eta \mathrm{eff}}$ obtained in sluggish lid regime solutions
presented in Section \ref{subsec_characterizatoinTR}
imply that they have thick boundary layers
and might become unstable due to the Rayleigh--Taylor instability. 
We performed time integration starting from some of the steady solutions
in order to examine their stability. 

The time-dependent equations are eqs.~\eqref{eq_energy} to \eqref{eq_BC0}.
Each time integration is started
from the given steady solution
adding a small pointwise disturbance with the amplitude of $T \times 10^{-3}$
for $T$ at the center of the domain. 
Time integrations are performed to $t=10$
with a time step of $10^{-6}$.
Steady solutions are qualified as stable if they evolve back to their initial steady state and unstable if they become time-dependent or reach a steady state that differs from the starting one.

\begin{figure}
  \centering
  \includegraphics[width=0.6\linewidth]{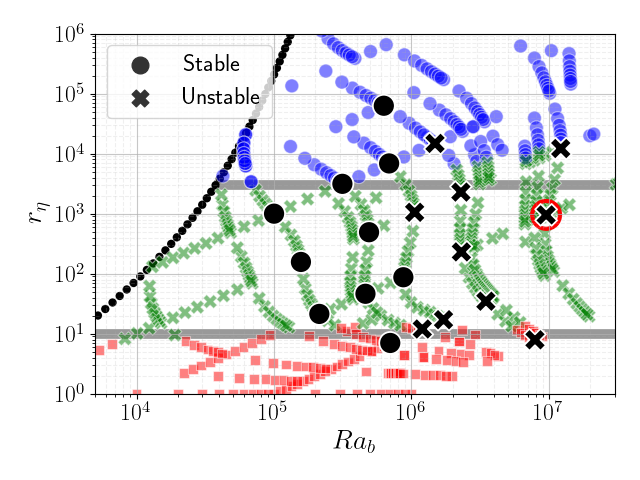}
  \caption{
    Stability of steady solutions. 
    Black circles and crosses express
    stable and unstable solutions, 
    respectively.
    A red circle identifies the solution which is shown in \figref{fig_plumeevolution}.
  }
  \label{fig_stabilityexamination}
\end{figure}
The results in terms of stability are presented in \figref{fig_stabilityexamination}.
Stable and unstable steady solutions are found in all regimes,
and their boundary is located at $Ra_b \approx 10^6$.
For all the cases of unstable solutions,
periodic growth of cold plumes from the active boundary layer is observed.

\begin{figure}
    \centering
    \begin{tabular}{cc} 
    \multirow{3}{*}{
        \includegraphics[width=0.6\linewidth]{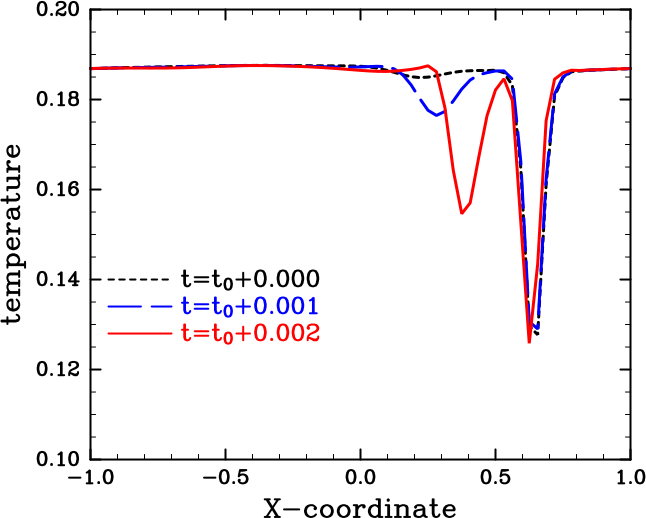}}
        &
      \includegraphics[width=0.35\textwidth]{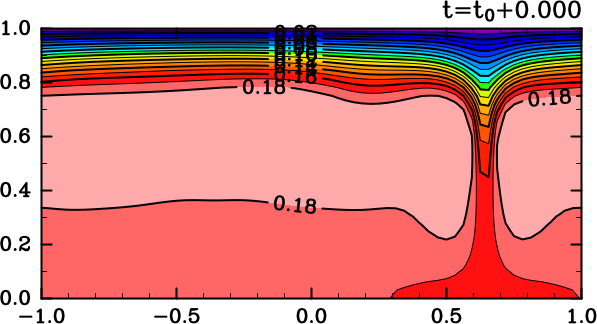}
        \\[0.2em]
      &\includegraphics[width=0.35\textwidth]{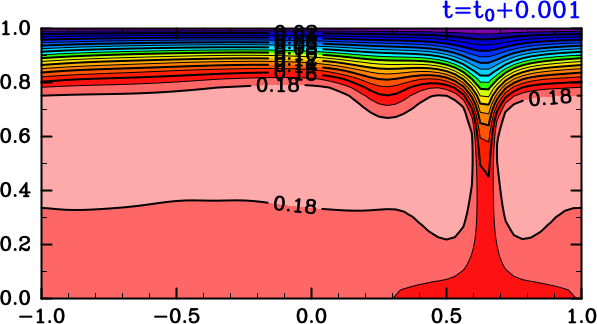}
      \\[0.2em]
      & \includegraphics[width=0.35\textwidth]{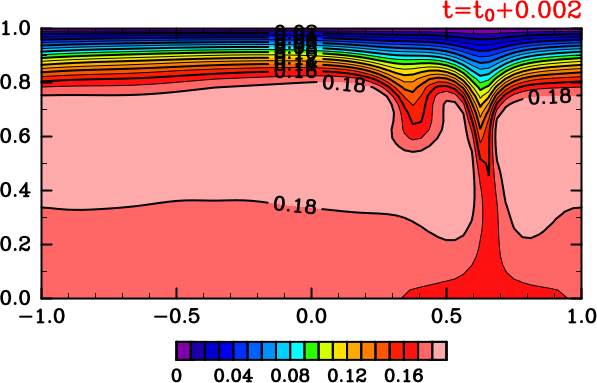}
    \end{tabular}
    \caption{
    Snapshots of temperature distributions obtained by time integration for $Ra_0 = 1 \times 10^4$ and $\gamma=40$. 
    The left panel shows time evolution of temperature 
    near the bottom of the boundary layer, at $y=0.68$.  
    $t$ is dimensionless time, and $t_0$ is the time of plume growth initiation.
    Two cold plumes can be observed; 
    the larger one corresponds to that of the steady solution
    and stays at $x=0.6$. 
    The smaller one grows due to the Rayleigh--Taylor instability and is carried from left to right by convective flow.
    The right panels show the evolution of temperature for the same time sequence as in the left panel. 
    }
    \label{fig_plumeevolution}
\end{figure}

A typical time evolution of the temperature field
of an unstable solution ($Ra_0 = 10^4$ and $\gamma=40$)
is shown in \figref{fig_plumeevolution}. 
It can be observed that
a small downward plume grows in the boundary layer
and is carried by the convective flow
to the location of the cold strong plume. 
Previous numerical models have found a similar behavior in which newly formed plumes are carried away by the horizontal flow generated by existing plumes. Merging of plumes leads to irregular patterns \citep[e.g.][]{Travis1994, Kameyama2000}.
\cite{Parmentier2000} attempt to estimate the
scaling of plume advection times and predict the number of
descending plumes in a region.
Following these observations, we assume that the limit between stability and instability is set by the relative speed of two processes: growth of a boundary layer instability and its entrainment by the preexisting flow. The mechanism for the growth of a plume instability in the boundary layer is Rayleigh--Taylor instability \citep[e.g.][]{Ribe2018}, which can be used to determine the growth rate of plume formation.
The time scale for the entrainment is set by the horizontal velocity of the steady solution, for which we presented a scaling law in the previous section.

\begin{figure}
  \centering
  \includegraphics[width=0.7\columnwidth]{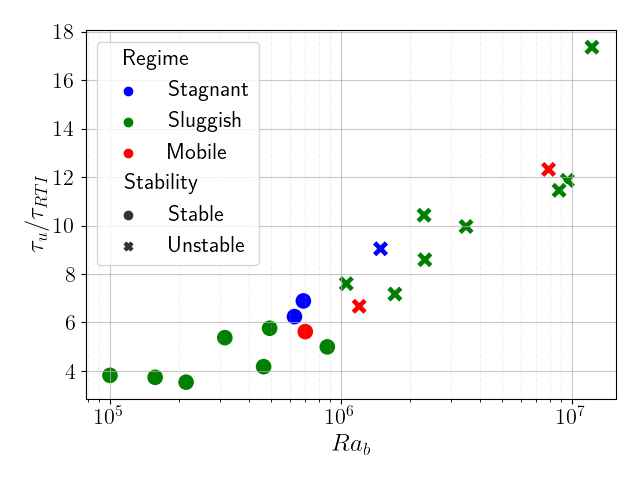}
  \caption{Results of time-dependent calculations on a diagram of $Ra_b$ and a ratio of time scales of horizontal entrainment to the Rayleigh--Taylor instability.
  Circles and crosses denote stable and unstable solutions, 
  respectively. 
  }
  \label{fig_timescaleratio}
\end{figure}

\figref{fig_timescaleratio} compares
these two time scales for the time integration solutions, and presents their importance for stability. 
The $y$ axis, $\tau_{u}/\tau_{\mathrm{RTI}}$, is a ratio of two time scales:
the time scale of horizontal entrainment by convective flow $\tau_{u}$, 
and the growing time scale of a Rayleigh--Taylor instability of the boundary layer between the isothermal core and the conductive lid $\tau_{\mathrm{RTI}}$:
\begin{equation}
    \tau_u = \frac{ \lambda }{ 2 u }, \quad
    \tau_{\mathrm{RTI}} = \left[
        Ra_b r_{\eta \mathrm{eff}}^{-1} \delta \Delta T_{\mathrm{eff}}
        f \left(r_{\eta \mathrm{eff}}, 2\pi \delta / \lambda \right) 
        \right]^{-1}.
    \label{eq_timescale}
\end{equation}
Here, $\Delta T_{\mathrm{eff}}$ is temperature-difference across the active boundary layer, and
$r_{\eta \mathrm{eff}} = \exp(\gamma \Delta T_{\mathrm{eff}})$ is the associated viscosity contrast.
$f(r_{\eta \mathrm{eff}}, 2\pi \delta / \lambda)$ is
the dimensionless growth-rate of the Rayleigh--Taylor instability derived by
\cite{whitehead1975} (their eq.~A15),
based on the linear stability analysis
of an interface between a thin fluid layer and a fluid in an infinite half-space with different density and viscosity.
Other factors in $\tau_{\mathrm{RTI}}$ of eq.~\eqref{eq_timescale} are for translation to the dimensionless time scale.
The growth-rate  $\tau_{\mathrm{RTI}}$ is estimated
by substituting the values of $r_{\eta \mathrm{eff}}$, $\delta$, and $\lambda$ of our convective solutions to the function $f$.

From \figref{fig_timescaleratio}, 
it can be understood that the stability of the steady solution
is determined by the ratio between the growing time scale of the Rayleigh--Taylor instability and the horizontal entrainment speed. 
The stability boundary is located at $\tau_u / \tau_{\mathrm{RTI}} \approx 6.67$, 
while
the stable and unstable solutions can be found for $Ra_b < 10^6$ and $Ra_b > 10^6$, respectively. 
Note that the stability of the solutions does not depend
on their regime. 
Therefore, sluggish lid regime solutions with large viscosity contrast
across the boundary layer can exist stably
for $Ra_b < 10^6$. 
Moreover, steady solutions with a thin strong downward plume
and broad slow upward flow (Section \ref{subsec_derivationofscaling})
exist between $10^5 < Ra_b < 10^6$ in the mobile regime.  

\subsection{Comparison with rigid surface cases}
\label{subsec_rigid}
In Section \ref{subsec_characterizatoinTR}, 
we showed that
the solutions in the sluggish lid regime have a larger viscosity contrast across the boundary layer
than that of the solutions in the stagnant lid regime.
To investigate the cause of this feature, 
we performed steady-state calculations with a rigid boundary condition on the top surface, 
where eq.~\eqref{eq_BC1} is replaced by
\begin{equation}
    T = \psi = \frac{ \partial \psi }{ \partial y } = 0,
    \quad 
    \text{at} ~ y=1.
    \label{eq_BC1rigid}
\end{equation}

\begin{figure}
  \centering
  \includegraphics[width=0.7\columnwidth]{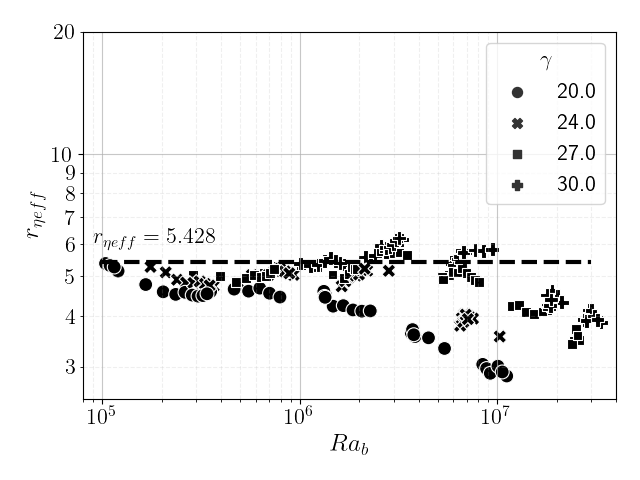}
  \caption{
  A plot of viscosity contrast $r_{\eta \mathrm{eff}}$ as a function of the Rayleigh number $Ra_b$ for steady solutions with the rigid upper boundary condition. 
  }
  \label{fig_retaeff-Rab_rigid}
\end{figure}

\figref{fig_retaeff-Rab_rigid} shows
a $Ra_b$--$r_{\eta \mathrm{eff}}$ plot
for the steady solutions with the rigid top boundary condition, 
which should be compared with \figref{fig_retaeff-Rab} 
for the stress-free condition. 
The regimes of the solutions cannot be distinguished by the mobility criterion used above,
because horizontal flow vanishes at the top surface due to the rigid condition there. 
Although the calculations are performed in the ranges of $Ra_0$ and $\gamma$ 
where the solutions of all regimes are obtained for the stress-free condition, 
no `high $r_{\eta \mathrm{eff}}$' solution is found.

\begin{figure}
    \centering
    \includegraphics[width=0.48\linewidth]{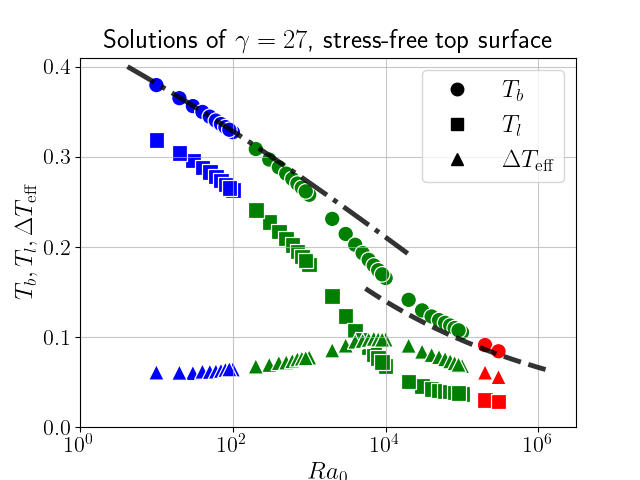}
    \includegraphics[width=0.48\linewidth]{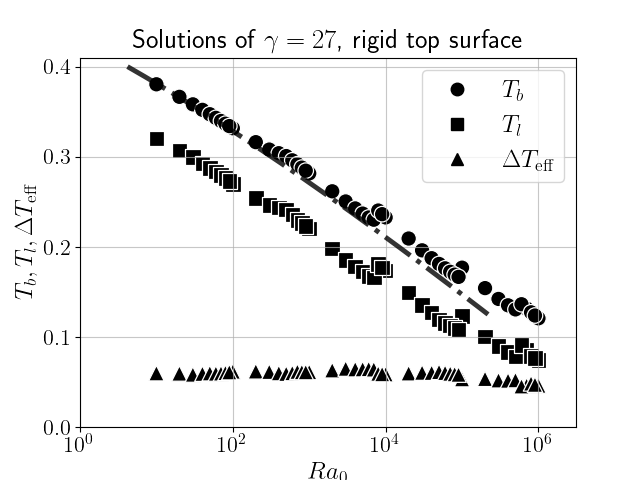}
    \caption{
    Dependencies of $T_b$ (circles), $T_l$ (squares), and $\Delta T_{\mathrm{eff}}=T_b - T_l$ (triangles) on $Ra_0$ are shown for the solutions of $\gamma=27$. 
    The cases with a stress-free top surface (left panel)  and a rigid top surface (right panel) are shown.
    Dash-dotted lines are the scaling function for the stagnant lid regime (eq.~\ref{eq_Ra0function_ST}) 
    and dashed line is that for the mobile regime (\ref{eq_Ra0function_SVC}). The color code on the left panel is the same as in the previous figures.
    }
    \label{fig_Temperature_freeandrigid}
\end{figure}

In order to further clarify the situation, we examine the behavior of the horizontally averaged bottom temperature $T_b$, the horizontally averaged temperature at the top of the active boundary layer $T_l$, and their difference $\Delta T_{\mathrm{eff}}=T_b - T_l$.
\figref{fig_Temperature_freeandrigid} shows the dependence of the temperatures on $Ra_0$ for the solutions of $\gamma=27$ with different boundary conditions.
In the left panel with the stress-free top boundary, the steep $T_b$ curve in the sluggish lid regime appears due to the large difference in the scaling of $T_b$ in the stagnant lid regime and the mobile lid regime.
The $T_l$ curve of the stagnant lid regime is almost parallel to the $T_b$ curve, resulting in a constant $\Delta T_{\mathrm{eff}}$, which is consistent with the findings of previous studies \citep[e.g.][]{Davaille1993, Grasset1998}.
Since $T_l$ in the mobile regime is close to 0 by definition,
the $T_l$ curve in the sluggish regime 
in a range $10^3 \lesssim Ra_0 \lesssim 10^4$ is steeper than that of $T_b$,
in order to reach $\approx 0$ for $Ra_0 \gtrsim 10^4$.
Then $\Delta T_{\mathrm{eff}}$ becomes larger than those of the stagnant lid regime and the mobile regime, resulting in
the prominent profile of $r_{\eta \mathrm{eff}}$ shown in \figref{fig_retaeff-Rab}.
On the other hand, on the right panel for the case of the rigid condition, 
a single scaling of $T_b$ as a function of $Ra_0$ seems to apply over the entire range.
Moreover, the $T_l$ curve is smooth and almost parallel to the $T_b$ curve.
As a result, $\Delta T_{\mathrm{eff}}$ is almost constant, resulting in the $r_{\eta \mathrm{eff}}$ plot without protrusion (\figref{fig_retaeff-Rab_rigid}).

The comparison of the solutions between the different top dynamical boundary conditions elucidates that the prominent $r_{\eta \mathrm{eff}}$ in the sluggish lid regime is due to the differences in the scaling relationship of $T_b$ between the stagnant lid regime and the mobile regime.
In the case of the stress-free upper boundary condition, the convective motion is affected by the stress-free condition in the mobile regime, where no lid emerges.
In the stagnant lid regime, the upper boundary of the active convective layer is adjacent to the conductive lid without motion, meaning that the effective dynamical condition is close to the rigid one.
Therefore, the large discrepancy of the scaling relations between the stagnant lid regime and the mobile regime can be attributed to the effective upper boundary condition.
In contrast, when the rigid condition is applied at the upper surface, the effective upper boundary condition for the active convective motion would not change drastically along the evolution of the conductive lid.

\subsection{Comparison to model of the Arrhenius viscosity}
\label{subsec_arrhenius}

We have adopted the Frank-Kamenetskii (FK) approximation viscosity for the internal heating convection problem discussed so far,
because it has been widely used for many numerical studies as a fundamental rheological model of mantle convection. 
However, it is useful to see the effects of different rheological models
for applications to planetary problems.
Here, we briefly perform numerical calculations of internal heating convection with the Arrhenius law viscosity and compare their results with those with the FK approximation viscosity.

The dimensionless form of the Arrhenius-law viscosity is given as \citep[e.g.][]{Grigne2023}
\begin{equation}
    \eta (T) = \exp\left[
                E_a \left( \frac{1}{T+T_o} - \frac{1}{T_o} \right)
                \right] 
             = \exp \left[ \frac{- E_a T}{ T_o (T+T_o)} \right],
    \label{eq_arrheniusvisc}
\end{equation}
where 
$E_a$ is dimensionless activation energy
and $T_o$ is temperature offset. 
$T_o$ can be seen as the dimensionless absolute temperature at which the viscosity is equal to 1, which is then the temperature that sets the viscosity entering the definition of the Rayleigh number $Ra_0$ (eq.~\ref{eq_Ra0}). However, as shown by \cite{Stein2013}, this temperature should be taken as the absolute temperature at the top surface divided by the temperature scale.
The variable corresponding to the FK parameter $\gamma$ in eq.~\eqref{eq_viscosity} can be written as 
\begin{equation}
    \gamma_A (T) = \frac{ E_a }{ T_o \left( T+T_o \right) } .
    \label{eq_arrheniusgamma}
\end{equation}
The FK approximation assumes that $\gamma$ is independent of temperature while $\gamma_A$ depends on the absolute temperature $T+T_o$.  

\begin{figure}
    \centering
    \begin{tabular}{cc}
    $T_o = 1.0$ & $T_o = 0.5$ \\
    \includegraphics[width=0.48\columnwidth]{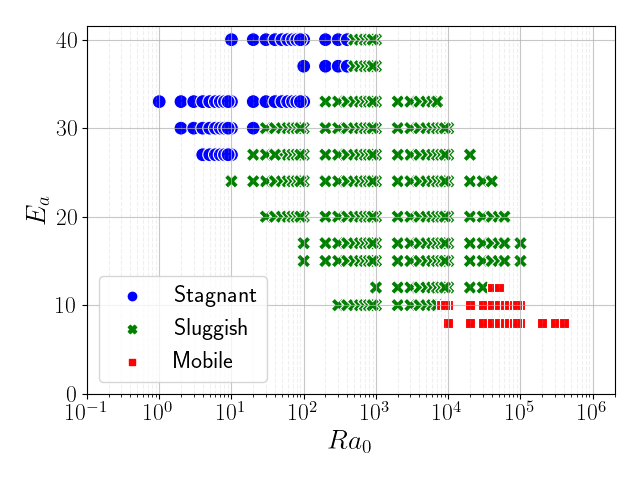} &
    \includegraphics[width=0.48\columnwidth]{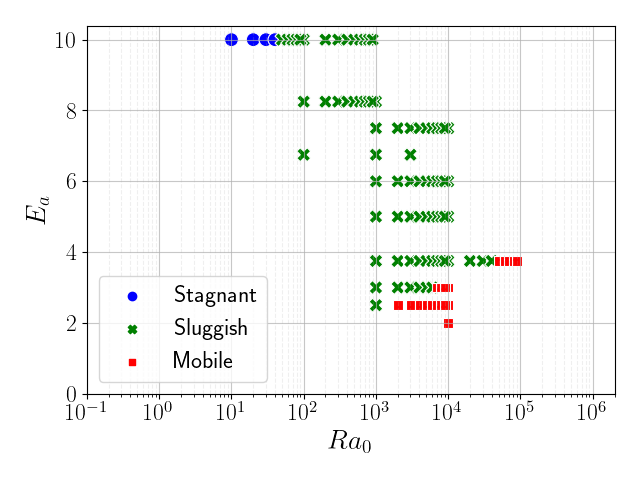} 
    \end{tabular}
    \caption{
    Steady solutions with Arrhenius viscosity on $Ra_0$--$E_a$ diagram. 
    Left and right panels are for $T_o=1.0$ and $T_o=0.5$, respectively. 
    Colors indicate the regimes of the solutions, meaning of which are the same as those in \figref{fig_regimediagram}. The regime classification is based on mobility described in Section \ref{subsec_steadysolutions}.
    }
    \label{fig_arrheniussolutions}
\end{figure}

We seek 
steady solutions with the Arrhenius viscosity \eqref{eq_arrheniusvisc} 
for various values of $E_a$ and $Ra_0$
using the Newton method described in Section \ref{subsec_newtonmethod}.
The first estimate for the iterative calculation is given by a FK viscosity solution whose value of $\gamma$ is equal to $\gamma_A (0)$.
Two values of the temperature offset $T_o = 1.0$ and $0.5$ are taken to evaluate the effect of the absolute temperature on the Arrhenius system.
\figref{fig_arrheniussolutions} shows the steady solutions obtained and the regime classification on $Ra_0$--$E_a$.
It can be seen that even when the Arrhenius viscosity is applied to the internal heating convection problem, three different regimes are identified similar to the ones obtained with the FK viscosity law.

\begin{figure}
    \centering
    \begin{tabular}{cc}
    \multicolumn{2}{c}{Arrhenius law ($T_o = 1.0$)} \\
    \includegraphics[width=0.48\columnwidth]{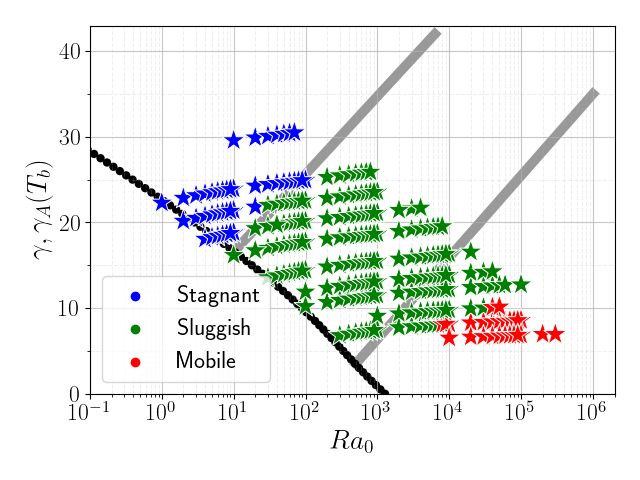} &
    \includegraphics[width=0.48\columnwidth]{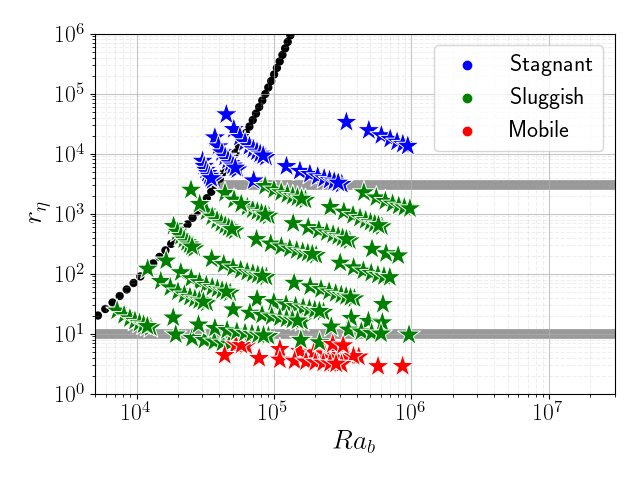} \\
    \multicolumn{2}{c}{Arrhenius law ($T_o = 0.5$)} \\
    \includegraphics[width=0.48\columnwidth]{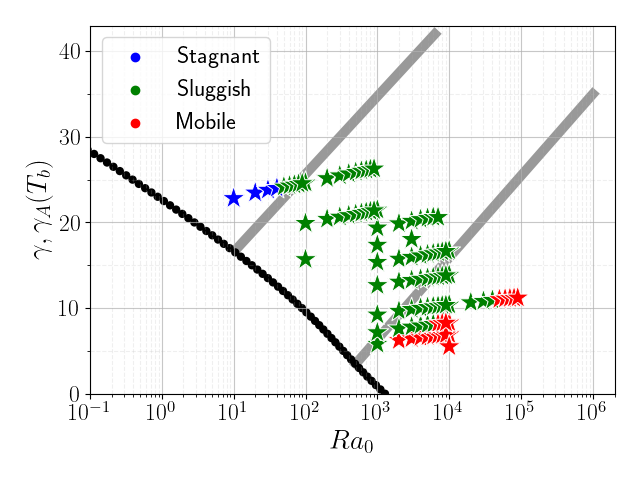} &
    \includegraphics[width=0.48\columnwidth]{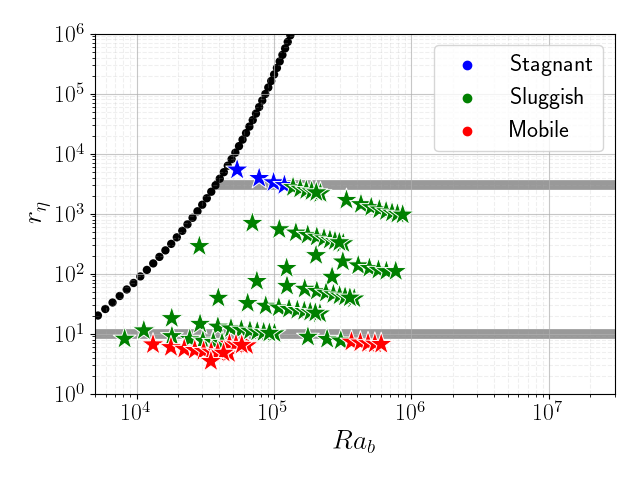} 
    \end{tabular}
    \caption{
    Regime diagrams for the Arrhenius viscosity solutions on the parameter planes with $Ra_0$--$\gamma_A(T_b)$ (left) and $Ra_b$--$r_\eta$ (right).
    The colors of each solution indicate the regimes distinguished in the same way as described in section \ref{subsec_steadysolutions}.
    Grey lines and black dots show the same regime boundaries and neutral curves for the FK viscosity cases indicated in \figref{fig_regimediagram}, where $\gamma$ is used in the vertical axis of the left panels instead of $\gamma_A(T_b)$.
    Upper and lower panels show the cases with $T_o=1.0$ and $T_o=0.5$, respectively.
    }
   \label{fig_regimediagram_arrhenius}
\end{figure}

\figref{fig_regimediagram_arrhenius} compares the regime diagrams of Arrhenius viscosity convection with those of FK viscosity convection shown in \figref{fig_regimediagram}.
By evaluating the values of $\gamma_A (T_b)$ from the solutions of the Arrhenius viscosity and projecting the solutions to the $Ra_0$--$\gamma_A (T_b)$ plane, we examine the consistency of the regime classification between Arrhenius viscosity convection and FK viscosity convection.
The same regime boundaries and neutral curve on each panel as shown in \figref{fig_regimediagram}
are represented by grey lines and black circles,
while $\gamma$ is used instead of $\gamma_A(T_b)$ on the vertical axis of the left panels.
The regime boundaries with the Arrhenius viscosity law agree fairly well with that obtained for the FK viscosity law, even though, for the mobile--sluggish lid boundary, it appears to be at a slightly lower value of $\gamma_A (T_b)$ (left panels). The agreement in the $Ra_b$--$r_\eta$ space (right panels) seems to be even better than in the $Ra_0$--$\gamma_A(T_b)$ space.

We also investigate the effect of the temperature offset $T_o$, the parameter of the Arrhenius viscosity law. 
Comparing the diagrams for $T_o=1.0$ (top panels) and $T_o=0.5$ (bottom panels) of
\figref{fig_regimediagram_arrhenius}, the sluggish--stagnant boundary is not influenced by the value of $T_o$ in that range, while the change of the mobile--sluggish boundary is not clear, which should be examined further by obtaining more steady solutions.
In addition, 
we should investigate the cases with smaller $T_o$, since the difference between the Arrhenius viscosity and the FK viscosity would become significant when $T_o < T$.
Indeed, looking at eq.~\eqref{eq_arrheniusgamma}, we see that for $T\ll T_o$, $\gamma_a(T)$ tends to a constant value, similar to the FK viscosity law. 
\cite{Grigne2023} showed that, 
in the case of bottom heating convection, 
different results are obtained for $T_o=0.2$, a value for the Earth's surface,and $T_o=1$, 
a value often used in previous studies. 
The situation is rendered more complex in the case of internal heating 
where the mean temperature decreases with the Rayleigh number, and thus $T_o$ should be taken less than $0.2$ when considering the Earth's condition. 
In our calculations, we typically have $T<T_o$, which limits the effects of the Arrhenius law and is far from the Earth's condition. More work will be needed to investigate this question systematically, in particular with lower values of $T_o$.

\section{Conclusions}
\label{sec_conclusion}
We have investigated the dynamical regimes of convection in a temperature-dependent fluid driven by homogeneous internal heating, 
with stress-free boundary conditions at both horizontal boundaries.
Finite amplitude steady solutions in a two-dimensional channel
of aspect ratio 2 were obtained
by the Newton method for several values of the Rayleigh number defined using the viscosity at the top boundary, $Ra_0$,
and the parameter, $\gamma$, defining the strength of the temperature-dependence of viscosity. 

By classifying the solutions with the top surface mobility, 
we found the sluggish lid regime between the mobile regime and the stagnant lid regime, 
which has not been clearly analyzed so far.
The solutions in the sluggish lid regime are characterized
by the thickness of the top conductive lid increasing with $Ra_0$
while the Rayleigh number defined with the viscosity at the bottom of the domain, $Ra_b$, remains nearly constant.
The viscosity contrast between the bottom of the conductive lid and the isothermal core of convection
is larger in the sluggish lid regime than in the other regimes.
When the stress-free boundary condition of the top surface is replaced by a rigid one,
the transition from the mobile regime to the stagnant lid regime is smooth, and clear and prominent intermediate solutions cannot be found.
This suggests that the sluggish lid regime is caused by the change in the effective dynamical boundary condition at the top of the active convective layer.
The dependences of the temperature at the bottom of the conductive lid $T_l$ and in the convective core $T_b$ on the Rayleigh number indicate that there are different scaling relations between the mobile regime and the stagnant lid regime in the cases with the stress-free surface boundary, which makes the prominent temperature difference and viscosity contrast across the active boundary layer, resulting in the sluggish lid regime.

We have derived scaling relationships
for the Nusselt number as a function of the Rayleigh number
defined using the bottom viscosity
for our steady solutions in the mobile regime and the stagnant lid regime
and compared them to those obtained by \cite{Grasset1998}. 
Our scaling laws at moderate Rayleigh number, $Nu \propto Ra^{1/4}$, 
agree with those of GP98,
but we find a lower exponent for large Rayleigh number solutions, $Nu \propto Ra^{1/6}$. 
To explain that scaling law, 
we have proposed a theory that relies on a velocity structure more focused on a strong down-welling plume obtained at high Rayleigh number.
The difference in the exponent for the higher Rayleigh number solutions in the mobile regime between our results and GP98 may be due to the time dependencies.
GP98's solutions are obtained by time integrations, and the aspect ratio of their convection cells varies with time, whereas those of our steady solutions are constant.
Since the aspect ratio could affect the scaling relationships, the exponent discrepancy would occur.
In fact, an exponent of $1/5$ has been presented for bottom heating convection \citep[e.g.][]{MorrisCanright1984, Fowler1985, MoresiSolomatov1995}, which is also smaller than $1/3$ of the classical Rayleigh--B\'enard convection.

We have investigated the stability of steady solutions
through time integrations. 
The results showed that
the steady solutions are stable when the Rayleigh number is small, $Ra_b < 10^6$,
while those for the large Rayleigh number become unstable
and temporally periodic features of the downward plumes are observed.
By using the Rayleigh--Taylor instability model,
the stability of convection system
can be explained quantitatively by comparing the timescales for plume growth and convective entrainment. 
The stability of the steady solutions is independent of their regime, 
i.e., sluggish-lid convection with a large viscosity contrast across the thermal boundary layer
can be stable in time-dependent simulations.

Finally, we have obtained steady solutions of internal heated convection with the Arrhenius-law viscosity. 
As in the cases of the Frank-Kamenetskii approximation, three regimes are identified using the method of regime classification by the mobility.
By converting the Arrhenius-law viscosity parameters into the FK viscosity parameter, it is found that the obtained regime diagrams for the Arrhenius-law viscosity solutions qualitatively agree with those for the Frank-Kamenetskii approximation, although there is a slight discrepancy in the location of the regime boundary between the mobile regime and the sluggish lid regime.
The regime diagram does not change significantly when the value of the temperature offset in the Arrhenius viscosity formula is reduced, which should be investigated more closely in the future by changing the temperature offset more drastically. 
The brief comparisons between Arrhenius viscosity convection and FK viscosity convection show that the regime diagrams obtained with the FK approximation qualitatively represent those of Arrhenius viscosity convection.
It is expected that our results obtained with the FK approximation would be useful for extending to the Arrhenius viscosity convection problem.

The regime diagrams obtained in our study will serve as fundamental knowledge
for studies on the surface environment and dynamics of the interior
of the inner and extrasolar terrestrial planets. 
It may be possible to estimate the dynamical states of the interior of terrestrial exoplanets
from observations of their surfaces realized in the future.
It may also be useful to deduce the evolution of the internal structure
over the thermal history of the planets. 
For example, using the results of a one-dimensional model and the regime diagram obtained in this study, we can predict the style of mantle convection
throughout the thermal evolution of planets.

Based on the present study,
we should investigate the effects of the horizontal scale of convection on the regime diagram and stability of steady solutions. 
The aspect ratio may affect the emergence and the different scaling laws for the symmetric cell type and the strong downward plume type described in Section \ref{subsec_derivationofscaling}.
Furthermore, it is interesting to examine
whether horizontally elongated solutions in the sluggish lid regime
may become preferable or not,
since horizontally elongated solutions have been found in steady convection driven by bottom heating
\citep{Okuda2023}. 
Very long wavelength solutions have also been obtained in time-dependent models of convection driven by internal heating \citep{Vilella2017} or secular cooling \citep{Morison2021}.

Moreover, for geophysical and planetary applications,
the present results should be extended to three-dimensional and spherical geometries,
and implement realistic rheological models. 
For example, 
it is suggested that the spherical geometry has a significant effect on the remeasurement boundaries for bottom heating cases \citep{JAVAHERI2024}.
We have examined the steady solutions with Arrhenius-law viscosity models, but focused only on comparing the regime diagrams for the two viscosity models.
The characteristics of the solutions in each regime, as shown in the present study for the FK approximation viscosity, should be investigated.
The historical strain-stress constitutive equation is also important for modeling
of plate tectonics at the planetary surfaces
\citep[e.g.,][]{Ogawa2003}.
The effects of compressibility may become important for mantle convection
of the super-Earth can be discussed by replacing the Boussinesq equations by those of an anelastic model or the fully compressible equations \citep{Ricard2022}. 

\section*{Acknowledgements}
We thank two anonymous reviewers for their constructive and useful comments that helped improve this paper.
The present study was supported by the Research Institute for Mathematical Sciences,
a Joint Usage/Research Center located in Kyoto University.
This work was also supported in part by
the Japan Society for the Promotion of Science
(KAKENHI Grant Nos. 21H01155, 23K20878, and 24K00694),
and by the Japan Science and Technology Agency through the SPRING program
(Grant No. JPMJSP2110).
The numerical calculations were performed on the computer systems of the Research Institute for Mathematical Sciences, Kyoto University.
The library `gtool5' \citep{gtool5}
was used for the input and output of numerical data.
The products of the Dennou-Ruby project (http://www.gfd-dennou.org/arch/ruby/)
were used to draw some figures.

\section*{Declaration of generative AI and AI-assisted technologies in the writing process}
While preparing this manuscript, the authors used `DeepL’ to improve English usage. After using this tool, the authors reviewed and edited the content as needed and took full responsibility for the manuscript's content.
\section*{Contributor roles}
\begin{description}
\item Hisashi Okuda: Conceptualization, Methodology, Software, Formal analyses, Investigation, Writing--Original draft preparation, Writing--Review and editing.
\item Shin-ichi Takehiro: Conceptualization, Methodology, Funding acquisition, Writing--Review and editing.
\item St\'ephane Labrosse: Conceptualization, Methodology, Writing--Review and editing.
\end{description}

\appendix

\bibliographystyle{plainnat}
\bibliography{ref.bib}

\begin{thebibliography}{38}
\providecommand{\natexlab}[1]{#1}
\providecommand{\url}[1]{\texttt{#1}}
\expandafter\ifx\csname urlstyle\endcsname\relax
  \providecommand{\doi}[1]{doi: #1}\else
  \providecommand{\doi}{doi: \begingroup \urlstyle{rm}\Url}\fi

\bibitem[Andrews(1972)]{Andrews1972}
D.~J. Andrews.
\newblock Numerical simulation of sea-floor spreading.
\newblock \emph{J. Geophys. Res.}, 77\penalty0 (32):\penalty0 6470--6481, 1972.
\newblock \doi{10.1029/JB077i032p06470}.

\bibitem[Bercovici et~al.(2000)Bercovici, Ricard, and Richards]{Bercovici2000}
D.~Bercovici, Y.~Ricard, and M.~A. Richards.
\newblock The relation between mantle dynamics and plate tectonics: A primer.
\newblock \emph{Geophysical Monograph-AGU}, 121:\penalty0 5--46, 2000.

\bibitem[Chandrasekhar(1961)]{Chandrasekhar}
S.~Chandrasekhar.
\newblock \emph{Hydrodynamic and hydromagnetic stability}.
\newblock Oxford university press, 1961.

\bibitem[Christensen(1984)]{Christensen1984}
U.~Christensen.
\newblock Convection with pressure- and temperature-dependent non-newtonian
  rheology.
\newblock \emph{Geophys. J. R. Astr. Soc.}, 77\penalty0 (2):\penalty0 343--384,
  1984.
\newblock \doi{10.1111/j.1365-246X.1984.tb01939.x}.

\bibitem[Davaille and Jaupart(1993)]{Davaille1993}
A.~Davaille and C.~Jaupart.
\newblock Transient high-rayleigh-number thermal convection with large
  viscosity variations.
\newblock \emph{J. Fluid Mech.}, 253:\penalty0 141–166, 1993.
\newblock \doi{10.1017/S0022112093001740}.

\bibitem[Fowler(1985)]{Fowler1985}
A.~C. Fowler.
\newblock Fast thermoviscous convection.
\newblock \emph{Stud. App. Math.}, 72\penalty0 (3):\penalty0 189--219, 1985.
\newblock \doi{10.1002/sapm1985723189}.

\bibitem[Fowler and O'Brien(1996)]{FowlerOBrien1996}
A.~C. Fowler and S.~B.~G. O'Brien.
\newblock A mechanism for episodic subduction on venus.
\newblock \emph{J. Geophys. Res. Planets}, 101\penalty0 (E2):\penalty0
  4755--4763, 1996.
\newblock \doi{10.1029/95JE03261}.

\bibitem[Grasset and Parmentier(1998)]{Grasset1998}
O.~Grasset and E.~M. Parmentier.
\newblock Thermal convection in a volumetrically heated, infinite prandtl
  number fluid with strongly temperature-dependent viscosity: Implications for
  planetary thermal evolution.
\newblock \emph{J. Geophys. Res. Solid Earth}, 103:\penalty0 18171--18181,
  1998.

\bibitem[Grign\'e(2023)]{Grigne2023}
C.~Grign\'e.
\newblock Stagnant-lid convection: comparison of viscosity laws and uniform
  scaling approach for temperature and heat flux prediction.
\newblock \emph{Geophys. J. Int.}, 235\penalty0 (3):\penalty0 2410--2429, 10
  2023.
\newblock ISSN 0956-540X.
\newblock \doi{10.1093/gji/ggad375}.

\bibitem[Hüttig and Breuer(2011)]{Huttig2011}
C.~Hüttig and D.~Breuer.
\newblock Regime classification and planform scaling for internally heated
  mantle convection.
\newblock \emph{Phys. Earth Planet. Inter.}, 186:\penalty0 111--124, 2011.
\newblock \doi{10.1016/j.pepi.2011.03.011}.

\bibitem[Ishiwatari et~al.(2012)Ishiwatari, Toyoda, Morikawa, Takehiro, Sasaki,
  Nishizawa, Odaka, Otobe, Takahashi, Nakajima, Horinouchi, Shiotani, Hayashi,
  and development group]{gtool5}
M.~Ishiwatari, E.~Toyoda, Y.~Morikawa, S.~Takehiro, Y.~Sasaki, S.~Nishizawa,
  M.~Odaka, N.~Otobe, Y.~O. Takahashi, K.~Nakajima, T.~Horinouchi, M.~Shiotani,
  Y.-Y. Hayashi, and Gtool development group.
\newblock "gtool5": a fortran90 library of input/output interfaces for
  self-descriptive multi-dimensional numerical data.
\newblock \emph{Geoscientific Model Development}, 5\penalty0 (2):\penalty0
  449--455, 2012.
\newblock \doi{10.5194/gmd-5-449-2012}.

\bibitem[Jain and Solomatov(2022)]{JainSolomatov2022}
C.~Jain and V.~S. Solomatov.
\newblock {Onset of convection in internally heated fluids with strongly
  temperature-dependent viscosity}.
\newblock \emph{Phys. Fluids}, 34:\penalty0 096604, 2022.
\newblock \doi{10.1063/5.0105170}.

\bibitem[Jaupart et~al.(2015)Jaupart, Labrosse, Lucazeau, and
  Mareschal]{Treatise_Jaupart2015}
C.~Jaupart, S.~Labrosse, F.~Lucazeau, and J.-C. Mareschal.
\newblock 7.06 - temperatures, heat, and energy in the mantle of the earth.
\newblock In Gerald Schubert, editor, \emph{Treatise on Geophysics}, pages
  223--270. Elsevier, Oxford, second edition, 2015.
\newblock ISBN 978-0-444-53803-1.
\newblock \doi{10.1016/B978-0-444-53802-4.00126-3}.

\bibitem[Javaheri et~al.(2024)Javaheri, Lowman, and Tackley]{JAVAHERI2024}
P.~Javaheri, J.~P. Lowman, and P.~J. Tackley.
\newblock Spherical geometry convection in a fluid with an arrhenius thermal
  viscosity dependence: The impact of core size and surface temperature on the
  scaling of stagnant-lid thickness and internal temperature.
\newblock \emph{Phys. Earth Planet. Inter.}, 349:\penalty0 107157, 2024.
\newblock \doi{10.1016/j.pepi.2024.107157}.

\bibitem[Kameyama and Ogawa(2000)]{Kameyama2000}
M.~Kameyama and M.~Ogawa.
\newblock Transitions in thermal convection with strongly temperature-dependent
  viscosity in a wide box.
\newblock \emph{Earth Planet. Sci. Lett.}, 180\penalty0 (3):\penalty0 355--367,
  2000.
\newblock \doi{10.1016/S0012-821X(00)00171-0}.

\bibitem[Labrosse et~al.(2018)Labrosse, Morison, Deguen, and
  {Alboussi\`ere}]{Labrosse_etal2018}
S.~Labrosse, A.~Morison, R.~Deguen, and T.~{Alboussi\`ere}.
\newblock Rayleigh-{B\'enard} convection in a creeping solid with a phase
  change at either or both horizontal boundaries.
\newblock \emph{J. Fluid Mech.}, 846:\penalty0 5--36, 2018.
\newblock \doi{doi.org/10.1017/jfm.2018.258}.

\bibitem[{McKenzie} et~al.(1974){McKenzie}, Roberts, and Weiss]{Mckenzie1974}
D.~P. {McKenzie}, J.~M. Roberts, and N.~O. Weiss.
\newblock Convection in the earth’s mantle: towards a numerical simulation.
\newblock \emph{J. Fluid Mech.}, 62\penalty0 (3):\penalty0 465–538, 1974.
\newblock \doi{10.1017/S0022112074000784}.

\bibitem[Moresi and Solomatov(1995)]{MoresiSolomatov1995}
L.‐N. Moresi and V.~S. Solomatov.
\newblock Numerical investigation of 2d convection with extremely large
  viscosity variations.
\newblock \emph{Phys. Fluids}, 7\penalty0 (9):\penalty0 2154--2162, 09 1995.
\newblock \doi{10.1063/1.868465}.

\bibitem[Morison et~al.(2021)Morison, Labrosse, and Choblet]{Morison2021}
A.~Morison, S.~Labrosse, and G.~Choblet.
\newblock Sublimation-driven convection in {Sputnik} {Planitia} on {Pluto}.
\newblock \emph{Nature}, 600:\penalty0 419--423, 2021.
\newblock \doi{10.1038/s41586-021-04095-w}.

\bibitem[Morris and Canright(1984)]{MorrisCanright1984}
S.~Morris and D.~Canright.
\newblock A boundary-layer analysis of benard convection in a fluid of strongly
  temperature-dependent viscosity.
\newblock \emph{Phys. Earth Planet. Inter.}, 36\penalty0 (3):\penalty0
  355--373, 1984.
\newblock \doi{10.1016/0031-9201(84)90057-8}.

\bibitem[Ogawa(2003)]{Ogawa2003}
M.~Ogawa.
\newblock Plate-like regime of a numerically modeled thermal convection in a
  fluid with temperature-, pressure-, and stress-history-dependent viscosity.
\newblock \emph{J. Geophys. Res. Solid Earth}, 108\penalty0 (B2), 2003.
\newblock \doi{10.1029/2000JB000069}.

\bibitem[Okuda and Takehiro(2023)]{Okuda2023}
H.~Okuda and S.~Takehiro.
\newblock Horizontal length of finite-amplitude thermal convection cells with
  temperature-dependent viscosity.
\newblock \emph{Phys. Earth Planet. Inter.}, 344:\penalty0 107103, 2023.
\newblock \doi{10.1016/j.pepi.2023.107103}.

\bibitem[Parmentier and Sotin(2000)]{Parmentier2000}
E.~M. Parmentier and C~Sotin.
\newblock Three-dimensional numerical experiments on thermal convection in a
  very viscous fluid: Implications for the dynamics of a thermal boundary layer
  at high rayleigh number.
\newblock \emph{Phys. Fluids}, 12\penalty0 (3):\penalty0 609--617, 2000.

\bibitem[Parmentier et~al.(1994)Parmentier, Sotin, and Travis]{Parmentier1994}
E.~M. Parmentier, C.~Sotin, and B.~J. Travis.
\newblock {Turbulent 3-D thermal convection in an infinite Prandtl number,
  volumetrically heated fluid: implications for mantle dynamics}.
\newblock \emph{Geophys. J. Int.}, 116\penalty0 (2):\penalty0 241--251, 02
  1994.
\newblock \doi{10.1111/j.1365-246X.1994.tb01795.x}.

\bibitem[Reese et~al.(2005)Reese, Solomatov, and Baumgardner]{Reese2005}
C.C. Reese, V.S. Solomatov, and J.R. Baumgardner.
\newblock Scaling laws for time-dependent stagnant lid convection in a
  spherical shell.
\newblock \emph{Phys. Earth Planet. Inter.}, 149\penalty0 (3):\penalty0
  361--370, 2005.
\newblock \doi{10.1016/j.pepi.2004.11.004}.

\bibitem[Ribe(2018)]{Ribe2018}
N.~M. Ribe.
\newblock \emph{Theoretical mantle dynamics}.
\newblock Cambridge University Press, 2018.

\bibitem[Ricard et~al.(2022)Ricard, Alboussi{\`{e}}re, Labrosse, Curbelo, and
  Dubuffet]{Ricard2022}
Y.~Ricard, T.~Alboussi{\`{e}}re, S.~Labrosse, J.~Curbelo, and F.~Dubuffet.
\newblock Fully compressible convection for planetary mantles.
\newblock \emph{Geophys. J. Int.}, 230\penalty0 (2):\penalty0 932--956, 2022.
\newblock \doi{10.1093/gji/ggac102}.

\bibitem[Sleep(2000)]{Sleep2000}
N.~H. Sleep.
\newblock Evolution of the mode of convection within terrestrial planets.
\newblock \emph{J. Geophys. Res. Planets}, 105\penalty0 (E7):\penalty0
  17563--17578, 2000.
\newblock \doi{doi.org/10.1029/2000JE001240}.

\bibitem[Solomatov(1995)]{Solomatov1995}
V.~S. Solomatov.
\newblock {Scaling of temperature‐ and stress‐dependent viscosity
  convection}.
\newblock \emph{Phys. Fluids}, 7\penalty0 (2):\penalty0 266--274, 02 1995.
\newblock \doi{10.1063/1.868624}.

\bibitem[Solomatov and Moresi(1996)]{SolomatovMoresi1996}
V.~S. Solomatov and L.-N. Moresi.
\newblock Stagnant lid convection on venus.
\newblock \emph{J. Geophys. Res. Planets}, 101\penalty0 (E2):\penalty0
  4737--4753, 1996.
\newblock \doi{10.1029/95JE03361}.

\bibitem[Solomatov and Moresi(2000)]{SolomatovMoresi2000}
V.~S. Solomatov and L.-N. Moresi.
\newblock Scaling of time-dependent stagnant lid convection: Application to
  small-scale convection on earth and other terrestrial planets.
\newblock \emph{J. Geophys. Res. Solid Earth}, 105\penalty0 (B9):\penalty0
  21795--21817, 2000.
\newblock \doi{10.1029/2000JB900197}.

\bibitem[Stein and Hansen(2013)]{Stein2013}
C.~Stein and U.~Hansen.
\newblock Arrhenius rheology versus frank-kamenetskii rheology—implications
  for mantle dynamics.
\newblock \emph{Geochem. Geophys. Geosyst.}, 14\penalty0 (8):\penalty0
  2757--2770, 2013.
\newblock \doi{10.1002/ggge.20158}.

\bibitem[Stengel et~al.(1982)Stengel, Oliver, and Booker]{Stengel1982}
K.~C. Stengel, D.~S. Oliver, and J.~R. Booker.
\newblock Onset of convection in a variable-viscosity fluid.
\newblock \emph{J. Fluid Mech.}, 120:\penalty0 411–431, 1982.
\newblock \doi{10.1017/S0022112082002821}.

\bibitem[Tackley(2000)]{Tackley2000}
P.~J. Tackley.
\newblock Self-consistent generation of tectonic plates in time-dependent,
  three-dimensional mantle convection simulations 2. strain weakening and
  asthenosphere.
\newblock \emph{Geochem. Geophys. Geosyst.}, 1\penalty0 (8), 2000.
\newblock \doi{10.1029/2000GC000043}.

\bibitem[Travis and Olson(1994)]{Travis1994}
B.~Travis and P.~Olson.
\newblock Convection with internal heat sources and thermal turbulence in the
  earth's mantle.
\newblock \emph{Geophys. J. Int.}, 118\penalty0 (1):\penalty0 1--19, 1994.

\bibitem[Turcotte and Schubert(2014)]{Geodynamics2014}
D.~Turcotte and G.~Schubert.
\newblock \emph{Geodynamics}.
\newblock Cambridge University Press, third edition, 2014.

\bibitem[Vilella and Deschamps(2017)]{Vilella2017}
K.~Vilella and F.~Deschamps.
\newblock Thermal convection as a possible mechanism for the origin of
  polygonal structures on {Pluto}'s surface.
\newblock \emph{J. Geophys. Res.}, 122:\penalty0 1056--1076, 2017.
\newblock \doi{10.1002/2016JE005215}.

\bibitem[Whitehead and Luther(1975)]{whitehead1975}
J.~A. Whitehead and D.~S. Luther.
\newblock Dynamics of laboratory diapir and plume models.
\newblock \emph{J. Geophys. Res.}, 80\penalty0 (5):\penalty0 705--717, 1975.
\newblock \doi{10.1029/JB080i005p00705}.

\end{thebibliography}

\end{document}